\documentclass[trackchanges,twocolumn]{aastex701}
\usepackage{amsmath}

\newcommand{\kms}{\,km\,s$^{-1}$} 

\newcommand{\hii}{H II}
\newcommand{\ha}{H${40\alpha}$}

\newcommand{\h}{H$_2$}

\newcommand{\meoh}{CH$_{3}$OH}
\newcommand{\ame}{CH$_{3}$OH-A v=0}
\newcommand{\eme}{CH$_{3}$OH-E v=0}
\newcommand{\vo}{v=0~}
\newcommand{\vt}{v$_{t}$=1}
\newcommand{\vtme}{CH$_{3}$OH v$_{t}$=1}
\newcommand{\vteme}{CH$_{3}$OH-E v$_{t}$=1}

\newcommand{\nh}{n$_{H_2}$}
\newcommand{\amax}{2(1,1)$-$1(1,0)$--$}
\newcommand{\emax}{12(-3,10)$-$13(-2,12)}
\newcommand{\ejj}{J(2,J-2)$-$J(-1,J-1)}
\newcommand{\vtmax}{6(1,6)$-$5(0,5)}

\newcommand{\vlsr}{V$_{\rm LSR}$}

\begin{document}

\title{The ALMA-ATOMS survey: Methanol emission in a large sample of hot molecular cores}

\correspondingauthor{Jiahang Zou; Tie Liu; Sheng-Li Qin}
\email{zojh1000@163.com; liutie@shao.ac.cn; qin@ynu.edu.cn}

\author[orcid=0009-0000-9090-9960]{Jiahang Zou}
\affiliation{School of Physics and Astronomy, Yunnan University, Kunming 650091, People's Republic of China}
\affiliation{State Key Laboratory of Radio Astronomy and Technology, Shanghai Astronomical Observatory, Chinese Academy of Sciences, \\
80 Nandan Road, Shanghai 200030, People's Republic of China}
\email{zojh1000@163.com}

\author[orcid=0000-0002-5286-2564]{Tie Liu} 
\affiliation{State Key Laboratory of Radio Astronomy and Technology, Shanghai Astronomical Observatory, Chinese Academy of Sciences, \\
80 Nandan Road, Shanghai 200030, People's Republic of China}
\email{liutie@shao.ac.cn}

\author[orcid=0000-0003-2302-0613]{Sheng-Li Qin}
\affiliation{School of Physics and Astronomy, Yunnan University, Kunming 650091, People's Republic of China}
\email{qin@ynu.edu.cn}

\author[orcid=0000-0001-5703-1420]{Yaping Peng}
\affiliation{Department of Physics, Faculty of Science, Kunming University of Science and Technology, Kunming 650500, People's Republic of China}
\email{pyp893@163.com}

\author[orcid=0000-0001-5950-1932]{Fengwei Xu}
\affiliation{Max Planck Institute for Astronomy, Königstuhl 17, 69117 Heidelberg, Germany}
\affiliation{Kavli Institute for Astronomy and Astrophysics, Peking University, 5 Yiheyuan Road, Haidian District, Beijing 100871, People's Republic of China}
\email{fengweilookuper@gmail.com}

\author[0000-0001-8315-4248]{Xunchuan Liu}
\affiliation{State Key Laboratory of Radio Astronomy and Technology, Shanghai Astronomical Observatory, Chinese Academy of Sciences, \\
80 Nandan Road, Shanghai 200030, People's Republic of China}
\email{liuxunchuan001@gmail.com}

\author[orcid=0009-0009-8154-4205]{Li Chen}
\affiliation{School of Physics and Astronomy, Yunnan University, Kunming 650091, People's Republic of China}
\email{li.chen@mail.ynu.edu.cn}

\author[orcid=0000-0002-4154-4309]{Xindi Tang}
\affiliation{State Key Laboratory of Radio Astronomy and Technology, Xinjiang Astronomical Observatory, Chinese Academy of Sciences, 830011 Urumqi, People's Republic of China}
\email{tangxindi@xao.ac.cn}

\author[orcid=0000-0002-8697-9808]{Sami Dib}
\affiliation{Max Planck Institute for Astronomy, K\"{o}nigstuhl 17, 69117, Heidelberg, Germany}
\email{sami.dib@gmail.com}

\author[0009-0005-7028-0735]{Zi-Yang Li}
\affiliation{School of Physics and Astronomy, Yunnan University, Kunming 650091, People's Republic of China}
\email{ziayngli@shao.ac.cn} 

\author[orcid=0000-0003-3343-9645]{Hong-Li Liu}
\affiliation{School of Physics and Astronomy, Yunnan University, Kunming 650091, People's Republic of China}
\email{hongliliu2012@gmail.com}

\author[orcid=0000-0002-5809-4834]{Mika Juvela}
\affiliation{Department of Physics, University of Helsinki, P.O. Box 64, FI-00014 University of Helsinki, Finland}
\email{mika.juvela@helsinki.fi}

\author[orcid=0000-0002-7125-7685]{Patricio Sanhueza}
\affiliation{Department of Astronomy, School of Science, The University of Tokyo, 7-3-1 Hongo, Bunkyo, Tokyo 113-0033, Japan}
\email{patosanhueza@gmail.com}

\author[orcid=0000-0002-8586-6721]{Pablo Garcia}
\affiliation{Chinese Academy of Sciences South America Center for Astronomy, National Astronomical Observatories, CAS, Beijing 100101, China}
\affiliation{Instituto de Astronom\'ia, Universidad Cat\'olica del Norte, Av. Angamos 0610, Antofagasta, Chile}
\email{astro.pablo.garcia@gmail.com}

\author[0000-0002-4154-4309]{Chang Won Lee}
\affiliation{University of Science and Technology, Korea (UST), 217 Gajeong-ro, Yuseong-gu, Daejeon 34113, Republic of Korea}
\affiliation{Korea Astronomy and Space Science Institute, 776 Daedeokdaero, Yuseong-gu, Daejeon 34055, Republic of Korea}
\email{cwl@kasi.re.kr}

\author[]{Guido Garay}
\affiliation{Chinese Academy of Sciences South America Center for Astronomy, National Astronomical Observatories, CAS, Beijing 100101, China}
\affiliation{Departamento de Astronomia, Universidad de Chile, Las Condes, 7591245 Santiago, Chile}
\email{guido@das.uchile.cl}

\author[0000-0001-7151-0882]{Swagat R. Das}
\affiliation{Departamento de Astronomia, Universidad de Chile, Las Condes, 7591245 Santiago, Chile}
\email{swagat@das.uchile.cl}

\author[orcid=0000-0001-7817-1975]{Yan-Kun Zhang}
\affiliation{State Key Laboratory of Radio Astronomy and Technology, Shanghai Astronomical Observatory, Chinese Academy of Sciences, \\
80 Nandan Road, Shanghai 200030, People's Republic of China}
\email{zhangyankun@shao.ac.cn}

\author[orcid=0000-0003-2412-7092]{Kee-Tae Kim}
\affiliation{University of Science and Technology, Korea (UST), 217 Gajeong-ro, Yuseong-gu, Daejeon 34113, Republic of Korea}
\affiliation{Korea Astronomy and Space Science Institute, 776 Daedeokdaero, Yuseong-gu, Daejeon 34055, Republic of Korea}
\email{ktkim@kasi.re.kr}

\author[orcid=0000-0003-3119-2087]{Jeong-Eun Lee}
\affiliation{Department of Physics and Astronomy, SNU Astronomy Research Center, Seoul National University, 1 Gwanak-ro, Gwanak-gu, Seoul 08826, Korea}
\email{lee.jeongeun@snu.ac.kr}

\author[orcid=0000-0002-5789-7504]{Meizhu Liu}
\affiliation{Center for Astrophysics, Guangzhou University, Guangzhou 510006, People's Republic of China}
\email{lmz@e.gzhu.edu.cn}

\author[]{Leonardo Bronfman}
\affiliation{Departamento de Astronom\'{i}a, Universidad de Chile, Las Condes, 7591245 Santiago, Chile}
\email{leo@das.uchile.cl}

\author[]{Zihping Kou}
\affiliation{State Key Laboratory of Radio Astronomy and Technology, Xinjiang Astronomical Observatory, Chinese Academy of Sciences, 830011 Urumqi, People's Republic of China}
\affiliation{University of Chinese Academy of Sciences, Beijing 100049, People’s Republic of China}
\email{kouzprl@gmail.com}

\author[]{Dongting Yang}
\affiliation{School of Physics and Astronomy, Yunnan University, Kunming 650091, People's Republic of China}
\email{dongting@mail.ynu.edu.cn}

\author[]{Gang Wu}
\affiliation{State Key Laboratory of Radio Astronomy and Technology, Xinjiang Astronomical Observatory, Chinese Academy of Sciences, 830011 Urumqi, People's Republic of China}
\email{wug@xao.ac.cn}

\author[0000-0001-7866-2686]{Jihye Hwang}
\affiliation{Institute for Advanced Study, Kyushu University, Japan}
\affiliation{Department of Earth and Planetary Sciences, Faculty of Science, Kyushu University, Nishi-ku, Fukuoka 819-0395, Japan}
\email{hwang.jihye.514@m.kyushu-u.ac.jp}

\author[0009-0000-5764-8527]{Dezhao Meng}
\affiliation{State Key Laboratory of Radio Astronomy and Technology, Xinjiang Astronomical Observatory, Chinese Academy of Sciences, 830011 Urumqi, People's Republic of China}
\affiliation{University of Chinese Academy of Sciences, Beijing 100049, People’s Republic of China}
\affiliation{State Key Laboratory of Radio Astronomy and Technology, Shanghai Astronomical Observatory, Chinese Academy of Sciences, \\
80 Nandan Road, Shanghai 200030, People's Republic of China}
\email{mengdezhao@xao.ac.cn}

\author[0000-0001-9160-2944]{Mengyao Tang}
\affiliation{Institute of Astrophysics, School of Physics and Electronical Science, Chuxiong Normal University, Chuxiong 675000, People's Republic of China}
\email{mengyao_tang@yeah.net}

\author[0000-0002-9875-7436]{James O. Chibueze}
\affiliation{UNISA Centre for Astrophysics and Space Sciences (UCASS),
College of Science, Engineering and Technology, University of South Africa, Cnr Christian de Wet Rd and Pioneer Avenue, Florida Park, 1709, Roodepoort, South Africa}
\affiliation{Department of Physics and Astronomy, University of Nigeria, 1 University Road, Nsukka 410001, Nigeria}
\email{james.chibueze@gmail.com}


\begin{abstract}

Methanol (\meoh) is a key complex organic molecule (COM) in the interstellar medium, widely used as a tracer of dense gas and hot molecular cores (HMCs). Using high-resolution ALMA observations from the ATOMS survey, we investigate the excitation and abundance of methanol nuclear spin isomers and their relationship to chemical complexity in massive star-forming cores. We identify 20 methanol transitions, including A- and E-type lines in the \vo state and E-type lines in the \vt\ state, and detect 94 HMC candidates. Rotational temperature analysis under the LTE assumption yields average values of 194 $\pm$ 33 K for \vteme, 178 $\pm$ 33 K for \ame, and 75 $\pm$ 33 K for \eme. Emission from COMs other than methanol is detected in 87 of the 94 cores, with the \vteme\ line intensity showing a strong correlation with the channel detection ratio (CDR). These results demonstrate that \vteme\ lines are reliable tracers of HMCs and chemical complexity, and that the CDR provides a robust indicator of molecular richness. The temperature difference between A- and E-type methanol transitions is driven by anomalously strong \ejj\ lines, highlighting the importance of analyzing methanol symmetry types separately. 

\end{abstract}



\section{Introduction}

Methanol (\meoh) is among the most abundant complex organic molecule (COM) in the interstellar medium (ISM), particularly in star-forming regions \citep{1970ApJ...162L.203B,1971ApJ...168L.101B,1986A&A...169..271M,2009ARA&A..47..427H}. It serves as a tracer of dense and warm gas \citep{1987ApJ...315..621B,1997ApJ...486..316B,2016A&A...595A.117J}, an astrochemical “clock” of ice-mantle processing, and a key parent species for the formation of larger COMs \citep{2008IAUS..251..143W,2014ApJ...788...68O}. Methanol forms primarily on icy dust-grain mantles via successive hydrogenation of CO and is released into the gas phase through thermal desorption in warm environments such as hot molecular cores and hot corinos \citep{2002ApJ...571L.173W,2006A&A...457..927G,2007A&A...474.1061B}. It exists as two nuclear-spin isomers (A- and E-types) with distinct symmetry states and statistical weights, which often lead to differences in excitation conditions and derived physical parameters \citep{1968JChPh..48.5299L,1988A&A...195..281F,2004A&A...422..573L}.

Hot molecular cores (HMCs) are compact ($n\gtrsim 10^6~ \text{cm}^{-3}$), warm ($T > 100~\text{K}$) gas structures associated with embedded high-mass protostars and are characterised by chemically rich spectra predominantly consist of COM emission \citep{2000prpl.conf..299K,2005IAUS..227...59C,2013A&A...559A..47B,2024A&A...686A.252B}. The advent of powerful interferometers — such as ALMA and NOEMA, and future arrays including ngVLA and SKA — has dramatically improved angular resolution and sensitivity, enabling access to weak COM transitions and resolving individual HMCs in clustered regions \citep{2005A&A...442..527M,2007A&A...474.1061B,2017A&A...604A..60B,2018ApJS..237....3S,2020A&A...635A.189F,2021ApJ...909..214L,2022MNRAS.511.3463Q,2022MNRAS.512.4419P,2025A&A...694A.166C,2025MNRAS.538.2579K}. Large sample surveys — for example, the ALMA-ATOMS/QUARKS survey \citep{Liu2020a,Liu2020b, 2021MNRAS.505.2801L,Liu2024RAA,2025ApJS..280...33Y} and the ALMAGAL survey \citep{Molinari2025} — now enable population-level studies of physical and chemical evolution of HMCs \citep{2025arXiv251101285M}. However, as spectral line datasets continue to grow rapidly, efficiently identifying genuine HMCs and characterising their chemical richness remains challenging. Full radiative-transfer modelling with tools such as WEEDS, CASSIS, XCLASS, MADCUBA and PYSPECKIT \citep{2011A&A...526A..47M,2015sf2a.conf..313V,2017A&A...598A...7M,2019A&A...631A.159M,2022AJ....163..291G} is excessively labour intensive and computationally demanding for large samples, motivating the development of simplified, physically motivated quantitative metrics for rapid and homogeneous assessment of molecular line richness and chemical diversity.

In this study, we used ALMA’s high resolution capabilities to conduct a detailed analysis of the A- and E-methanol lines. Our sample consists of 94 hot core candidates identified from ATOMS surveys \citep{2021MNRAS.505.2801L,2022MNRAS.511.3463Q,2022MNRAS.512.4419P,2025A&A...697A.190L,2025MNRAS.538.2579K}. We characterize the rotational temperatures and column densities to gain insight into its formation and excitation conditions. We also investigated the impact of non-LTE effects in environments where the hydrogen density falls below the critical thermalization threshold. Additionally, we assess the potential of methanol as a tracer for molecular line richness in both externally heated and internally heated dense cores, particularly in ultracompact H II regions.

The structure of this paper is as follows: Section \ref{obs} introduces the observations, Section \ref{sec:result} details our data processing methods and analysis, Section \ref{discussion} discusses our results and findings, and Section \ref{conclution} presents our conclusions.


\section{Observations and data reduction} \label{obs}

We utilized data from the ATOMS survey (Project ID: 2019.1.00685.S; Principal Investigator: Tie Liu\cite[][]{Liu2020a,Liu2020b}). The observations were conducted from September to mid-November 2019, targeting 146 IRAS clumps using both the Atacama Compact 7-m Array (ACA; Morita Array) and the 12-m array (in C43-2 or C43-3 configurations) in Band 3. The sample selection and basic observational parameters are detailed in \cite{Liu2020a,Liu2020b}. The correlator setup included eight Spectral Windows (SPWs): six with higher spectral resolutions of approximately $0.2 \text{--} 0.4 \, \text{km} \, \text{s}^{-1}$ (SPWs 1–6) in the lower sideband, and SPWs 7 and 8 with lower spectral resolutions in the upper sideband. SPWs 7 and 8 have a broad bandwidth of $1875 \, \text{MHz}$, corresponding to a spectral resolution of approximately $1.6 \, \text{km} \, \text{s}^{-1}$, and were used for continuum imaging and line surveys. The frequency ranges for SPWs 7 and 8 are $97536 \, \text{MHz}$ to $99442 \, \text{MHz}$, and $99470 \, \text{MHz}$ to $101390 \, \text{MHz}$, respectively, covering many COM lines. Due to the frequency setup, the methanol transition at 97582.8 MHz and at 97678.3 MHz was not included in SPW 7 for 18 sources. In addition to the COM lines, the \ha~line at $99023 \, \text{MHz}$ in SPW 7 is used to identify the \hii\ region , while the SiO (2-1) line at $86847 \, \text{MHz}$ in SPW 4 is used to trace the shocked gas. Since hot cores are compact and less affected by missing flux, we used only the 12-m array data to identify COM lines and hot cores, thereby minimizing contamination from the extended environment.

Data reduction was performed using the CASA software package (version 5.6; \citealt{2022PASP..134k4501C}). The resulting continuum images and line cubes for the 146 clumps have angular resolutions of approximately 1.2$''{-}$ 1.9$''$, corresponding to linear resolutions of $\sim$0.01-0.1 pc at the target distances (1.25${–}$11.5 kpc), and maximum recoverable angular scales of approximately 14.5$''{-}$ 20.3$''$. The mean $1\sigma$ noise level is better than $10 \, \text{mJy} \, \text{beam}^{-1}$ per channel for lines and $0.4 \, \text{mJy} \, \text{beam}^{-1}$ for the continuum. Considering an angular resolution of 1.9$''$ and a $3\sigma$ level, the positional accuracy of the line images due to noise is estimated to be better than 0.3$''$, using the formula $\Delta\theta = 0.45\frac{\theta_{FWHM}}{\text{SNR}}$ \citep{1988ApJ...330..809R}.

\section{Results} \label{sec:result}

\subsection{Methanol Emission Line in hot core candidates}
\label{sec 3.1}

We first extracted spectral lines from SPWs 7 and 8 for three representative hot cores: two located in I16272-4837 (c1 and c3; also known as SDC335; \citealt{2023MNRAS.520.3259X,2025ApJ...995..111Z}) and one in I18089-1732. All detected lines in these two windows were identified with the extended CASA line analysis software suite (XCLASS; \cite{2017A&A...598A...7M}; see Section \ref{sec 3.2} for details). Example identifications are presented in Fig \ref{fig:16272c1} in the Appendix \ref{app:C}. Among the secured line assignments, 20 transitions are unambiguously attributed to methanol and are listed in Table \ref{Table 1}. Among these, the lines at 97678 MHz, 97856 MHz and 99374 MHz correspond to a pair of transitions with similar energy levels. The \vteme~lines at 99772.8 and 99776.8 MHz are indistinguishable when the velocity width is larger than 3.5 \kms, while the \vteme~line at 99374.6 MHz blends with the HCCNC; v$_5$=1(10-9) \citep{1992ApJ...386L..51K,2024A&A...682L..13C}. 


\begin{table*}
\caption{Observed methanol transitions in SPWs 7 and 8.}
\label{Table 1}
\centering
\begin{tabular}{ccccc}
\hline\hline
Rest Frequency & Quantum Numbers & $E_u/k$ & Einstein $A$ & Detection \\
(MHz)          & $J(K_a,K_c)$    & (K)     & (s$^{-1}$)   & Counts    \\
\hline
\multicolumn{5}{c}{\ame}                                                               \\
 97\,582.798 &  2(1,1)$-$1(1,0)$--$        &   21.56 & $2.63\times10^{-6}$ & 84 \\
 97\,677.684 & 21(6,16)$-$22(5,17)$--$     &  729.27 & $1.44\times10^{-6}$ & 83 \\
 97\,678.803 & 21(6,15)$-$22(5,18)++       &  729.27 & $1.44\times10^{-6}$ & 83 \\
 97\,856.519 & 30(8,22)$-$31(7,25)++       & 1393.04 & $1.56\times10^{-6}$ & 20 \\
 97\,856.548 & 30(8,23)$-$31(7,24)++       & 1393.04 & $1.56\times10^{-6}$ & 20 \\
 99\,601.940 & 15(1,14)$-$15(1,15)$-+$     &  290.48 & $2.87\times10^{-8}$ & 24 \\
100\,585.006 & 28(1,27)$-$27(4,24)$--$     &  665.16 & $2.78\times10^{-8}$ & 12 \\
\hline
\multicolumn{5}{c}{\vteme}                                       \\
 99\,001.161 & 20(5,16)$-$21(6,16)          & 1019.57 & $2.70\times10^{-6}$ & 35 \\
 99\,374.341 & 15(6,10)$-$14(7,8)           &  771.01 & $1.02\times10^{-6}$ & 35 \\
 99\,374.624 & 15($-6$,10)$-$14($-7$,8)     &  771.01 & $1.02\times10^{-6}$ & 35 \\
 99\,730.940 & 6(1,6)$-$5(0,5)              &  335.31 & $2.79\times10^{-6}$ & 94 \\
 99\,772.834 & 20($-3$,17)$-$21($-4$,18)    &  897.53 & $4.97\times10^{-6}$ & 93 \\
 99\,776.775 & 20(3,18)$-$21(4,18)          &  897.53 & $4.98\times10^{-6}$ & 93 \\
\hline
\multicolumn{5}{c}{\eme}                                               \\
 98\,267.833 & 30(5,26)$-$29(6,23)          & 1201.58 & $1.70\times10^{-6}$ &  6 \\
100\,638.872 & 12($-3$,10)$-$13($-2$,12)    &  233.61 & $1.69\times10^{-6}$ & 94 \\
101\,097.069 & 3(2,1)$-$3($-1$,2)\textsuperscript{\dag}           &   39.83 & $1.30\times10^{-9}$ &  6 \\
101\,101.735 & 4(2,2)$-$4($-1$,3)\textsuperscript{\dag}           &   49.12 & $3.98\times10^{-9}$ & 22 \\
101\,126.857 & 5(2,3)$-$5($-1$,4)\textsuperscript{\dag}           &   60.72 & $9.58\times10^{-9}$ & 42 \\
101\,185.453 & 6(2,4)$-$6($-1$,5)\textsuperscript{\dag}           &   74.66 & $1.99\times10^{-8}$ & 70 \\
101\,293.415 & 7(2,5)$-$7($-1$,6)\textsuperscript{\dag}           &   90.91 & $3.73\times10^{-8}$ & 87 \\
\hline
\end{tabular}
\tablecomments{
\textsuperscript{\dag} Transitions J(2,J–2)–J(–1,J–1) with J=3–7 are anomalously bright (see Section~\ref{sec 4.1}), likely causing the rotational temperature to be underestimated and the column density overestimated.}

\end{table*}

\begin{figure*}[htbp]
\centering
\digitalasset
\begin{minipage}{\textwidth}
  \centering
  \includegraphics[page=15,width=\textwidth]{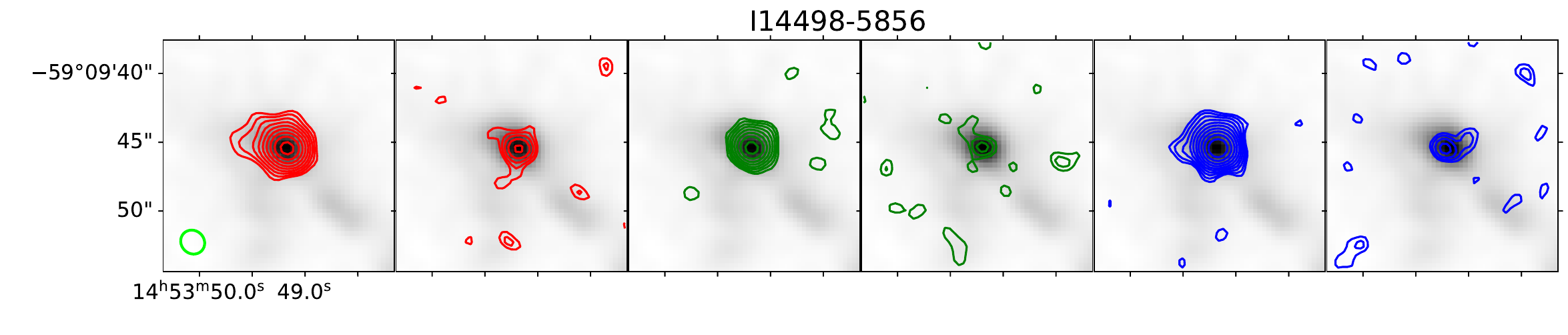}
\end{minipage}\\[1ex]
\begin{minipage}{\textwidth}
  \centering
  \includegraphics[page=5,width=\textwidth]{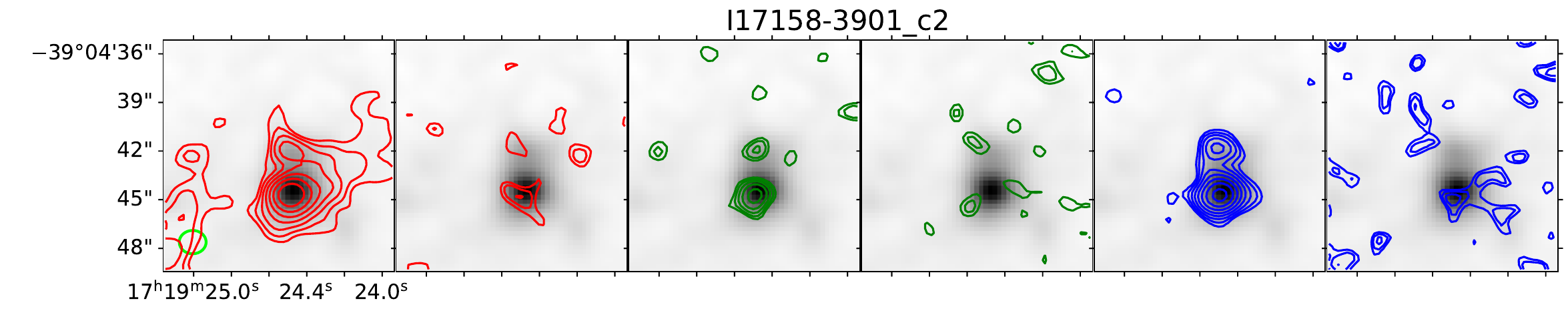}
\end{minipage}\\[1ex]
\begin{minipage}{\textwidth}
  \centering
  \includegraphics[page=43,width=\textwidth]{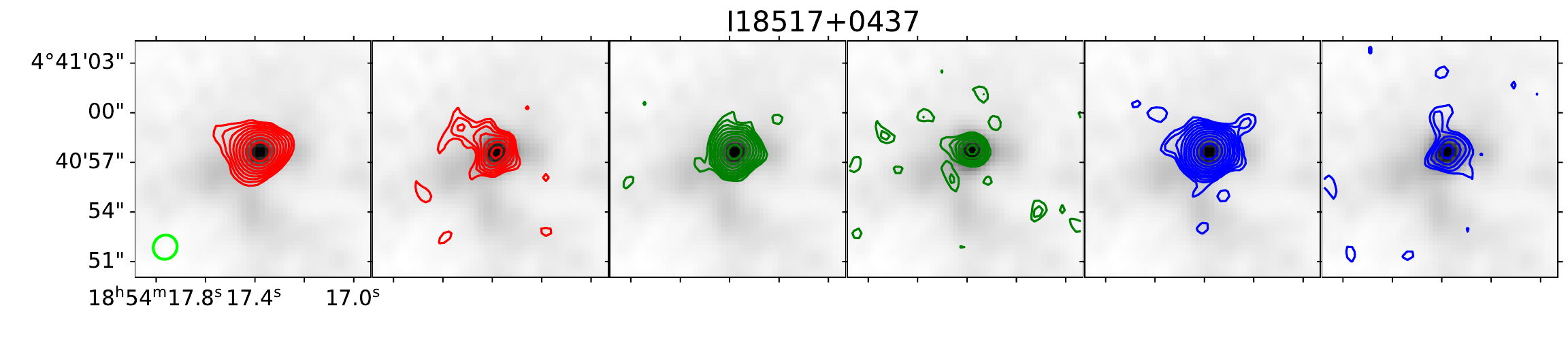}
\end{minipage}
\caption{Line Images. \textbf{Background:} 3-mm continuum (100 GHz), bottom left circle shows the beam. \textbf{Contours from left to right:} Red:\ame~at 97582.8 MHz, \ame~at 97678.3 MHz; Green: \vteme~at 99730.9 MHz, \vteme~at 99772 MHz and 99776 MHz; Blue: \eme~at 100638.9 MHz, \eme~at 101293.3 MHz. The E$_u$ of the transitions are indicated in the first figure, see Table \ref{Table 1}. The lowest contour level is set at 3$\sigma$, with each subsequent contour increasing by a factor of 1.4 times. The complete figure set (94 images) is available in the online journal.}
\label{Fig 1. lineimages}
\figsetstart
\figsetnum{1}
\figsettitle{Line images}

\figsetend

\end{figure*}

To facilitate spatial identification of hot core candidates, we created line images of selected \meoh~transitions, as described below. The systemic velocities of the cores (\vlsr) are adopted from either \cite{Liu2020b} or \cite{2022MNRAS.511.3463Q}. Based on \vlsr, we calculated the Doppler shifted frequencies for seven \meoh~transitions: 97582.8 MHz and 97678.3 MHz for \ame; 100638.9 MHz and 101293.3 MHz for \eme; 99730.9 MHz, 99772.8 and 99776.8 MHz for \vteme. Line images were generated from the moment 0 map within a velocity range of ±7 \kms, above a 5$\sigma$ detection limit, as shown in Fig.~\ref{Fig 1. lineimages}. The two transitions at 99772.8 and 99776.8 MHz are very close to each other and thus are integrated together. Cores exhibiting compact and strong \vtme~emission are classified as hot core candidates. The positions of the hot core candidates were determined based on the peak positions of methanol lines, in conjunction with the continuum emission, which was used to exclude spatially unassociated or extended emission. \citep{2011A&A...529A..24Z,2023ApJ...956...43L,2024ApJ...964...34L}.
Their coordinates are listed in Table \ref{Table 2}.


We then extracted and analyzed the spectral lines of 94 hot core candidates that exhibit compact emission in the strongest transition at 99730.9 MHz (\vteme; \vtmax). The detection counts of methanol transitions are provided in the last column of Table~\ref{Table 1}. The detection status of each transition frequency across these hot cores is summarized in Appendix Table~\ref{Table 3}. For 10 sources, three major transitions of methanol-A were not covered due to the spectral window setup issue described in Section~\ref{obs}, resulting in fewer detected lines and preventing subsequent fitting analyzes. The spectral lines and fitting results are shown in Fig.~\ref{Fig 2}, and will be further discussed in Section \ref{sec 3.2}. 87 show additional COM emission lines, including CH$_3$OCHO and C$_2$H$_5$CN, while 60 have been confirmed firmly as hot cores in \citet{2022MNRAS.511.3463Q}. The positions of these 60 hot cores were originally adopted from the 3 mm continuum core catalog of \citet{2021MNRAS.505.2801L}. However, the continuum emission in some sources is affected by free-free contamination. To improve positional accuracy, we revised the source sizes and coordinates of these 60 hot cores using both the 3 mm continuum and methanol line emission. Among the 60 cores, 55 show good agreement with the positions reported in \cite{2022MNRAS.511.3463Q}, while the remaining five were re-centered in this work based on their methanol emission peaks.

For the remaining 34 newly identified hot core candidates, we compared their spectral properties with those of the hot cores previously confirmed in \cite{2022MNRAS.511.3463Q}. Among them, 15 sources show strong COM lines and are preliminarily identified as hot cores, while 12 exhibit multiple weak ($\sim$0.5–1 K) COM transitions and are considered hot core candidates. The remaining 7 show no COM lines other than methanol above the 3$\sigma$ level and are therefore not classified as hot cores (marked with superscript “x” in in Table.~\ref{Table 3}).

\subsection{The line excitation of \meoh}
\label{sec 3.2}

Due to the threefold hindering potential, methanol is divided into A- and E-type symmetry species whose electric-dipole transitions are strictly forbidden between the two manifolds; each symmetry therefore possesses its own rotational energy ladder and spectrum \citep{1970ApJ...162L.203B,1988A&A...198..253M,2011A&A...533A..24W,2023ApJS..266...29Z,2005IAUS..231P..99L}. Previous studies have shown that forcing both symmetries into a single component often leads to poor simultaneous fits, particularly for optically thick or highly excited transitions \citep[e.g.,][]{2022MNRAS.511.3463Q,2022MNRAS.512.4419P}. We therefore treat A- and E-methanol as independent species throughout the analysis.

Under the LTE assumption, synthetic spectra are generated with XCLASS using six free parameters (source size, beam size, linewidth, velocity offset, rotational temperature, and column density), while transition frequencies, level degeneracies, and partition functions are drawn from the Cologne Database for Molecular Spectroscopy (CDMS; \citealt{2001A&A...370L..49M,2005JMoSt.742..215M,2016JMoSp.327...95E}) via the XCLASS database interface \citep{2017A&A...598A...7M}. The A- and E-type transitions are fitted independently as separate species. The statistical consistency between transition degeneracies and symmetry-specific partition functions, as well as the treatment of the intrinsic degeneracy of the E species, are discussed in detail in Appendix~\ref{app:C}.


\begin{figure*}[htbp]
\centering
\digitalasset
\begin{minipage}{\textwidth}
  \centering
  \includegraphics[page=5,width=\textwidth]{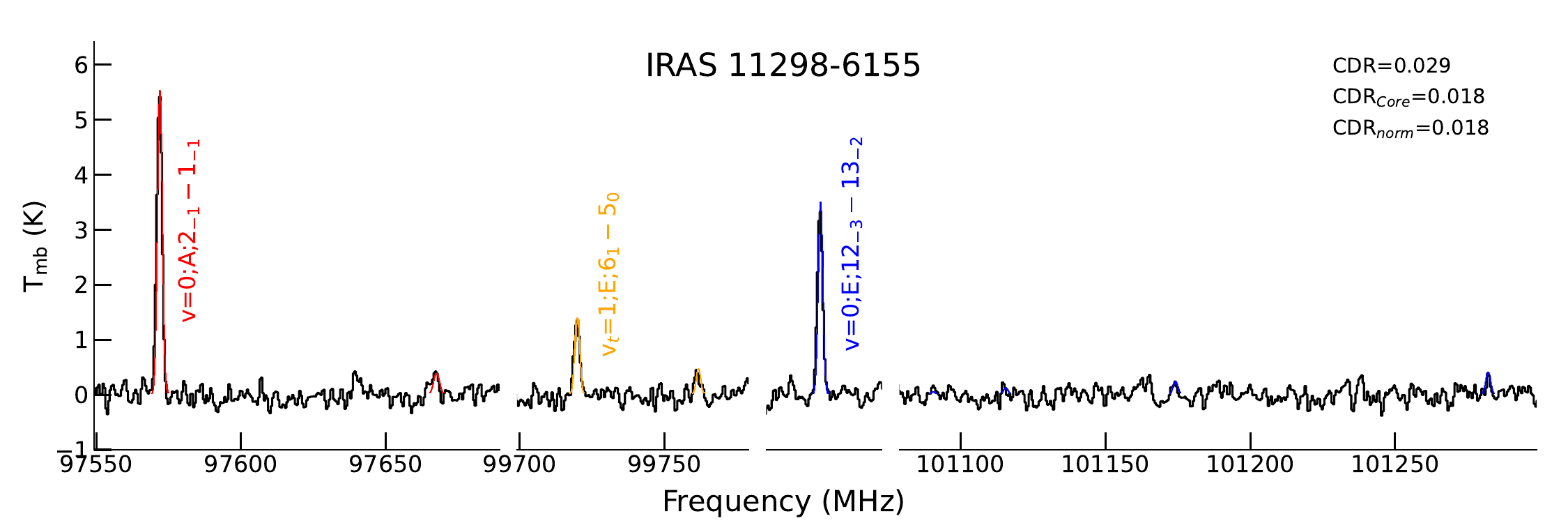}
\end{minipage}\\[1ex]
\begin{minipage}{\textwidth}
  \centering
  \includegraphics[page=24,width=\textwidth]{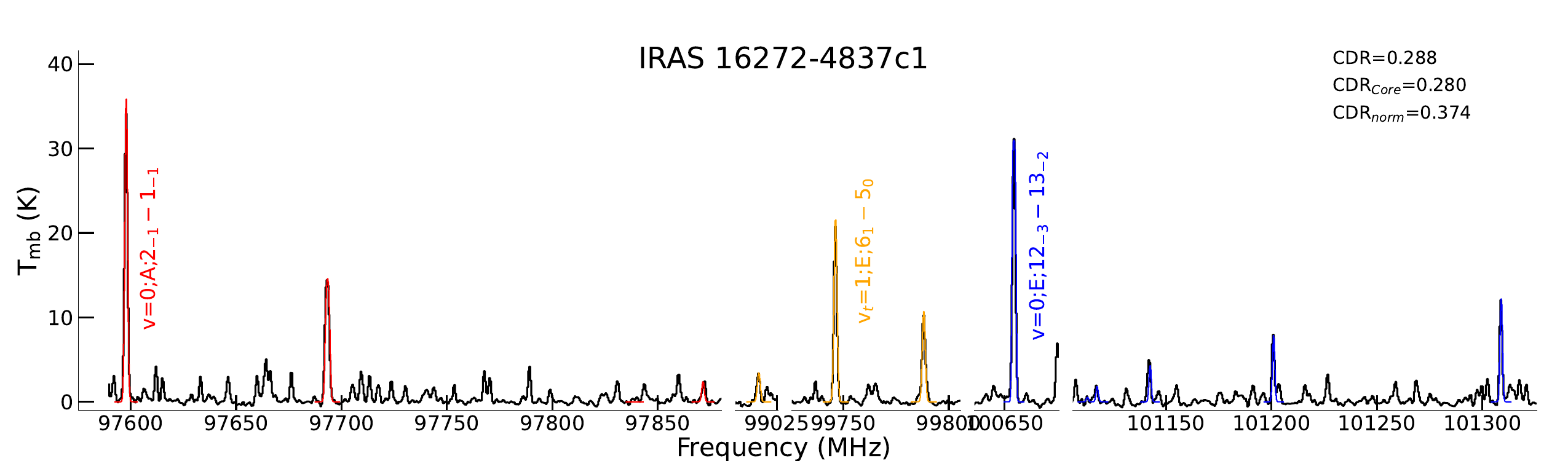}
\end{minipage}\\[1ex]
\begin{minipage}{\textwidth}
  \centering
  \includegraphics[page=44,width=\textwidth]{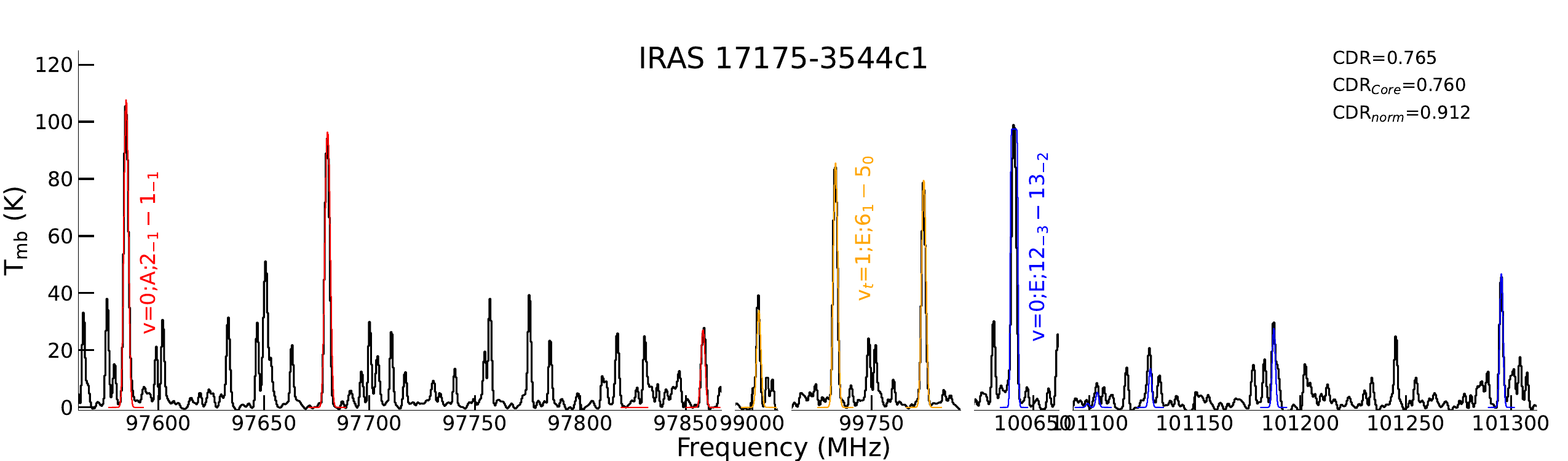}
\end{minipage}
\caption{Examples of the best-fit results from the myXCLASS function for spectra. The black line represents the observed values; the red, yellow, and blue lines represent the fitted values of \ame, \vteme, and \eme, respectively.  \textbf{Notes}: Hot Core I11298-6155 has no \ha~line and weak other molecular lines; hot core I16272-4837c1 has no \ha~line but shows rich emission lines of COMs; hot core/UC H II region I18507+0110 has a \ha~line and exhibits rich emission lines of COMs. The complete figure set (94 images) is available in the online journal.}
\label{Fig 2}
\figsetstart
\figsetnum{2}
\figsettitle{Spectra and the best-fit results.}

\figsetend
\end{figure*}


\begin{figure}[htbp]
\centering

\begin{minipage}{0.48\textwidth}
  \centering
  \includegraphics[width=\textwidth]{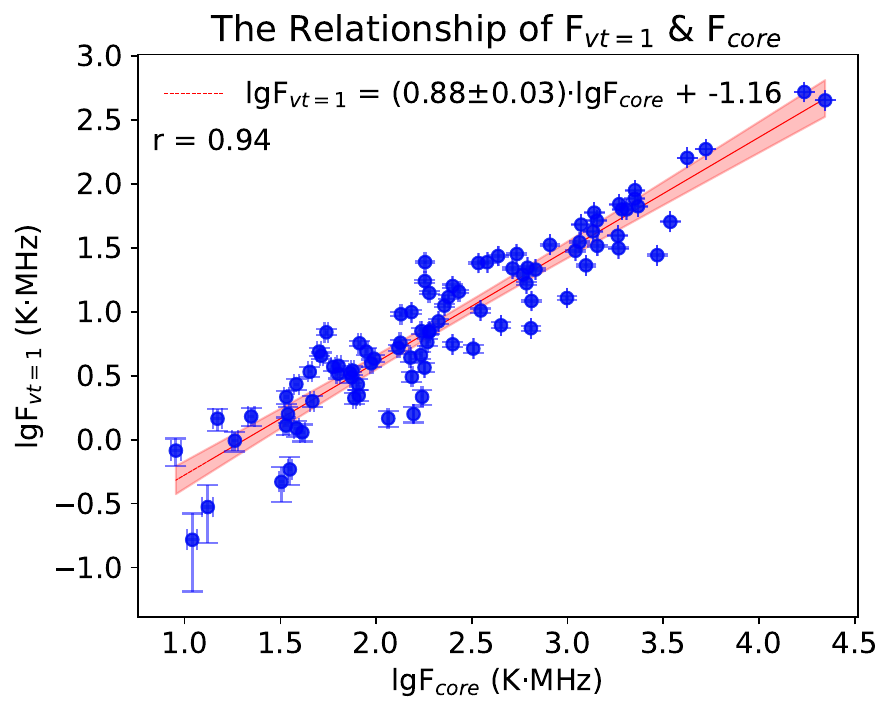}
\end{minipage}%
\hfill

\begin{minipage}{0.48\textwidth}
  \centering
  \includegraphics[width=\textwidth]{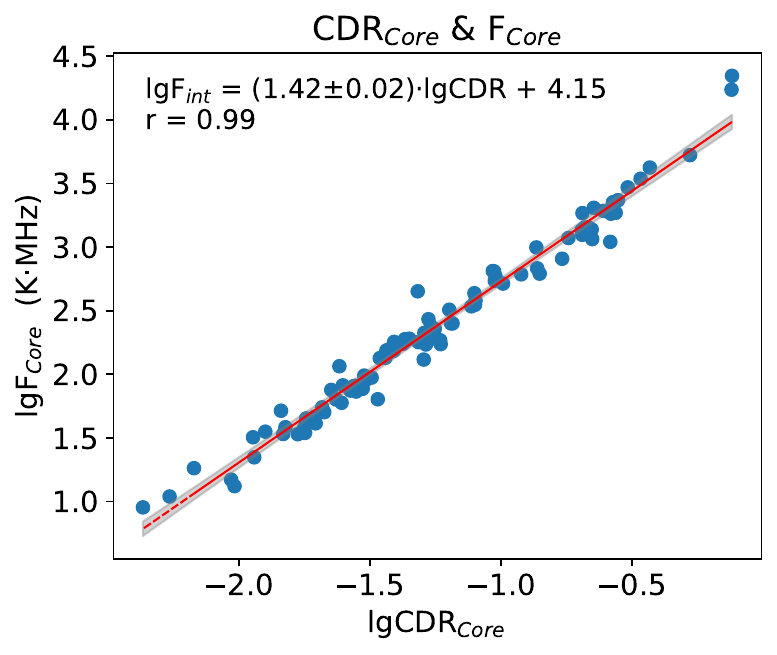}
\end{minipage}
\caption{\textbf{Top:} Integrated intensity of \vteme~vs. Integrated intensity of COMs in SPW7 and 8. \textbf{Bottom:} Integrated intensity vs. channel detection ratio of COMs.\\
\textbf{Note:}In calculating the integrated intensities, we excluded the following frequencies and their adjacent channels: 97981.0 MHz (CS); 99299.9 MHz (SO); 100076.4 MHz (HC$_3$N); 97714.9 MHz, 97702.3 MHz (SO$_2$); 99023.0 MHz (\ha); and other ionized emission lines at 99225 MHz, 98199 MHz, 98671 MHz, and 100540 MHz. We did not specifically exclude the H$_2$CCO, H$_2$CO, NH$_2$D, and high-vibration transitions of HC$_3$N\citep{2025A&A...694A.166C}, although these molecules are not COMs, their emission features are similar to those of COMs.}
\label{Fig 3}
\end{figure}

For the 66 cores without \ha~lines, we adopted the deconvolved continuum size as the source size. For 28 cores where the continuum was affected by free–free emission, we instead used the deconvolved size of the \eme (100638.9 MHz, \emax) transition, which was found to most closely match the continuum emission sizes in unaffected sources. Using these source sizes, we optimized the line parameters with XCLASS and refined the resulting rotational temperatures and column densities using the Model-Analysis Generic Interface for eXternal numerical codes \cite[MAGIX;][]{2013ascl.soft03009M} to estimate uncertainties. The resulting parameters are summarized in Table~\ref{Table 2}, and the corresponding spectra and fit results are shown in Fig.~\ref{Fig 2} and Appendix~\ref{app:C}.

During spectral fitting, each species involves five free parameters. Allowing all of them to vary in the MCMC optimization often led to poor convergence, especially when both the velocity and linewidth were free. We therefore constrained these two kinematic parameters during the line identification stage by manually adjusting them to reproduce the observed spectra. Fixing them at these values yields excellent fits while significantly improving the stability and efficiency of the MCMC runs, and this strategy was adopted for the entire sample.

\subsection{Quantitative Analysis of Methanol Emission and Molecular Line richness in Cores}
\label{sec 3.3}

Although hot molecular cores are well known for exhibiting chemically rich spectra predominated by COM emission, the degree of “richness” itself has remained only loosely defined, making quantitative comparisons between sources or large samples difficult. Previous efforts, such as counting the number of detected emission lines \citep{2020A&A...635A.198B}, have provided useful empirical indicators and were later adopted in the ATOMS and QUARKS surveys  \citep{2021MNRAS.505.2801L,2025ApJS..280...33Y}. However, the number of detectable molecular lines is an imperfect measure of richness: it can be dominated by a few bright species and becomes unreliable in crowded bands where blending suppresses line identifications. Therefore, a more physically motivated tracer is needed.

In this study, we first used vibrationally excited methanol (\vt) lines as tracers of the warmest gas and quantify richness through two integrated measures: the \vteme\ intensity ($F_{vt=1}$) and the total integrated intensity of molecules only originating in the hot core emission regions ($F_{core}$). We compute $F_{vt=1}$ using all transitions in Table~\ref{Table 1} except the 99374.6 MHz line, and derive $F_{core}$ by integrating channels above 3$\sigma$ while excluding transitions not confined to the hot core such as CS, SO, HC$_3$N 11–10, \ame;\amax, CH$_3$CHO $5_{1,4}$–$4_{1,3}$ \citep[e.g.,][]{2025ApJ...995..111Z}, SO$_2$, \ha, and other radio-recombination lines. A tight correlation is found between the two measures ($r=0.94$), given by $\log F_{vt=1} = 0.88 \log F_{core} - 1.16$. Cores showing COM emission (CH$_3$OCHO or C$_2$H$_5$CN) consistently exceed 35 K·MHz in $F_{core}$, whereas those below this threshold rarely display additional lines.

To further quantify molecular detectability, we introduce the Channel Detection Ratio (CDR):
\begin{equation}
\mathrm{CDR} = 
\frac{N_{\mathrm{sig} > 3\sigma}}{N_{\mathrm{chan}}},
\end{equation}

For consistency with $F_{int}$, we compute CDR only over the same set of usable channels and refer to this quantity as CDR$_{\mathrm{Core}}$. The lower pannel of Fig.~\ref{Fig 3} shows that CDR$_{Core}$ scales positively with the integrated intensity of molecular lines. The CDR$_{Core}$ for all cores is summarized in Table~\ref{Table 3} and visualized in Fig.~\ref{Fig 2} and Appendix~\ref{app:C}. Cores with strong \vtme\ emission typically show a higher CDR$_{Core}$, indicating richer molecular environments. The implications of these variations CDR are discussed in Section~\ref{sec 4.2}.

\section{Discussion} \label{discussion}

\begin{figure}[htbp]
\centering

\begin{minipage}{0.48\textwidth}
  \centering
  \includegraphics[width=\textwidth]{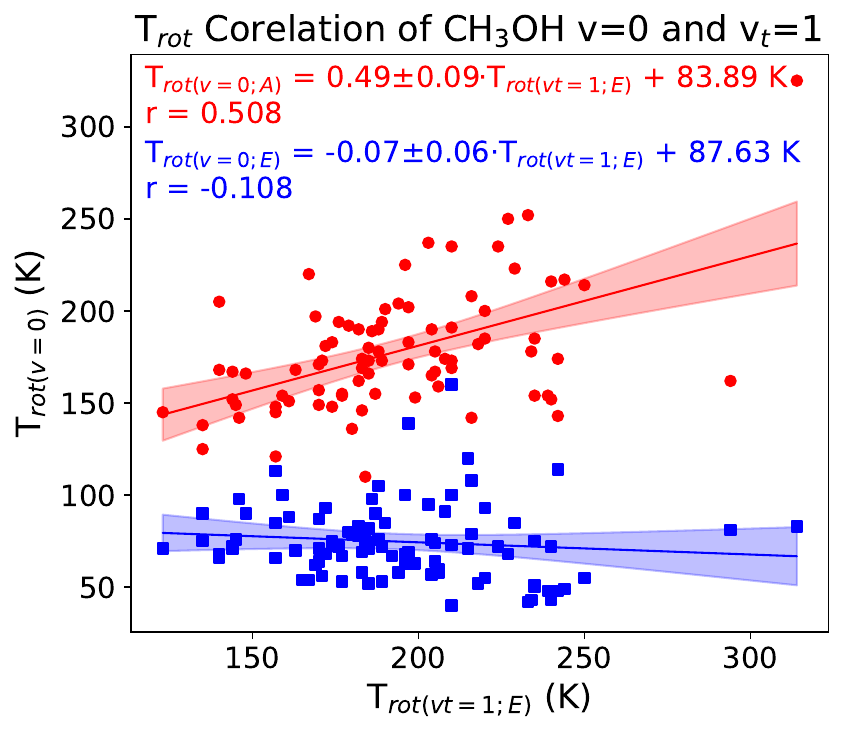}
\end{minipage}%
\hfill

\begin{minipage}{0.48\textwidth}
  \centering
  \includegraphics[width=\textwidth]{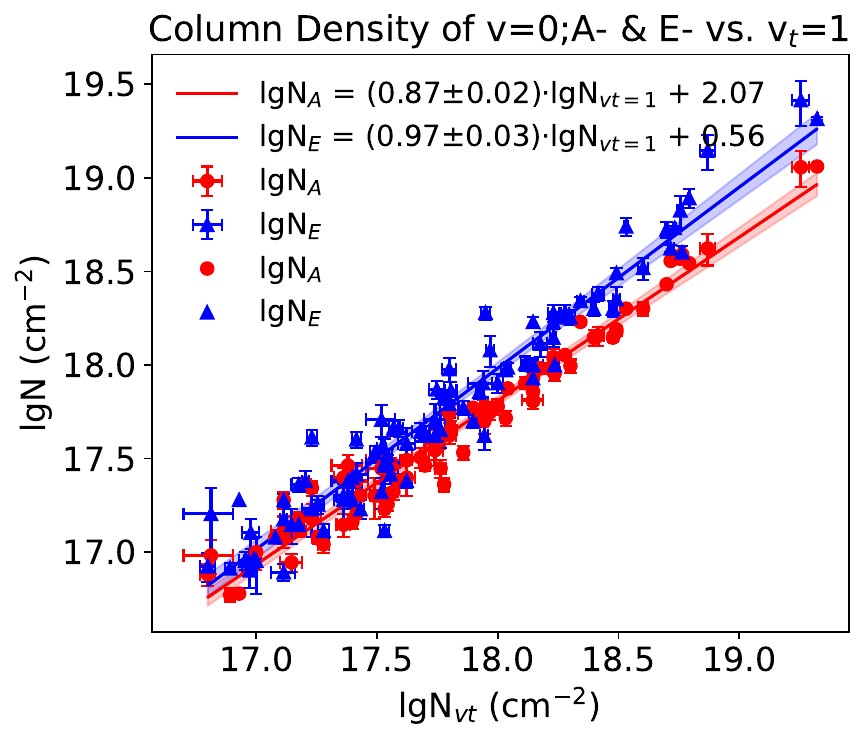}
\end{minipage}
\caption{
\textbf{Top:} The rotational temperature of \ame~(red) and \eme~(blue) versus \vteme. \textbf{Bottom:} The column density comparison for \ame~(red), \eme~(blue) with respect to \vteme. }
\label{Fig 4}
\end{figure}

\subsection{The Necessity of Separating \meoh\ Symmetry Types and Vibrational States in Hot Cores}
\label{sec 4.1}

Our spectral fitting reveal significant differences in excitation behavior between the A- and E-type \meoh~transitions within the \vo state. Fig.~\ref{Fig 4} compares their rotational temperatures (upper panel) and column densities (lower panel) with those of the vibrationally excited \vtme~lines: the \ame~transitions yield a mean rotational temperature of 178 $\pm$ 33 K, similar to that of \vtme~(194 $\pm$ 33 K), and show a moderate positive correlation between the two (\textit{r}=0.508). In contrast, the \eme~transitions give a much lower average temperature of 74 $\pm$ 21 K and a weak negative correlation (\textit{r}=--0.108). Under the LTE assumption commonly adopted for hot core analyses this temperature discrepancy is highly anomalous and drives the derived E-type column densities to values 54\% higher than those of A-type, which is chemically implausible.

Fitting the E-type spectrum with the same parameters used for A-type, we find that the low-energy $E_{2\rightarrow 1}$ transitions are anomalously strong: their observed intensities exceed the XCLASS LTE predictions by factors of 2--10 \citep[see Fig.~1 of][]{2022MNRAS.511.3463Q}. Their spatial distribution coincides with that of high-excitation lines, indicating that they originate in the hot, innermost regions of the cores rather than in cooler envelope material. To investigate this behavior, we performed RADEX modelling of these transitions. As shown in Fig.~\ref{fig:radex} in Appendi.~\ref{app:C}, the \ejj~lines approach LTE intensities only at densities above $n_{\mathrm{H}_2} \sim 3 \times 10^{9}~\mathrm{cm^{-3}}$, far exceeding the typical core-average values of $\sim 10^{7}~\mathrm{cm^{-3}}$ (Fig.~\ref{Fig 5}). Even at $n_{\mathrm{H}_2} = 6 \times 10^{8}~\mathrm{cm^{-3}}$, RADEX still predicts intensities well above LTE. Although its treatment of optical depth is simplified, the modelling clearly indicates that these levels remain strongly sub-thermal in most hot cores, causing the E-type rotational temperatures to be underestimated and their column densities to be overestimated.

Vibrationally excited \vtme~lines avoid these pitfalls. Their critical densities match the bulk of the hot core gas ($T > 100~\mathrm{K}$, $n_{\mathrm{H}_2} \sim 10^{7}~\mathrm{cm^{-3}}$; \citealt{1986A&A...169..271M,2010A&A...521L..39K,2022A&A...664A.171V}), their emission is spatially compact, and their integrated intensities correlate tightly with those of other complex molecules in hot cores ($r \approx 0.94$; Fig.~\ref{Fig 3}). We therefore recommend treating methanol symmetry types and vibrational states separately in hot core models: the ground state E-type transitions analyzed here yield apparent column densities that are systematically overestimated because of non-LTE excitation, whereas \vt~lines provides a robust probe of warm and dense regions.


\begin{figure*}[ht!]
\centering
\includegraphics[width=0.48\textwidth]{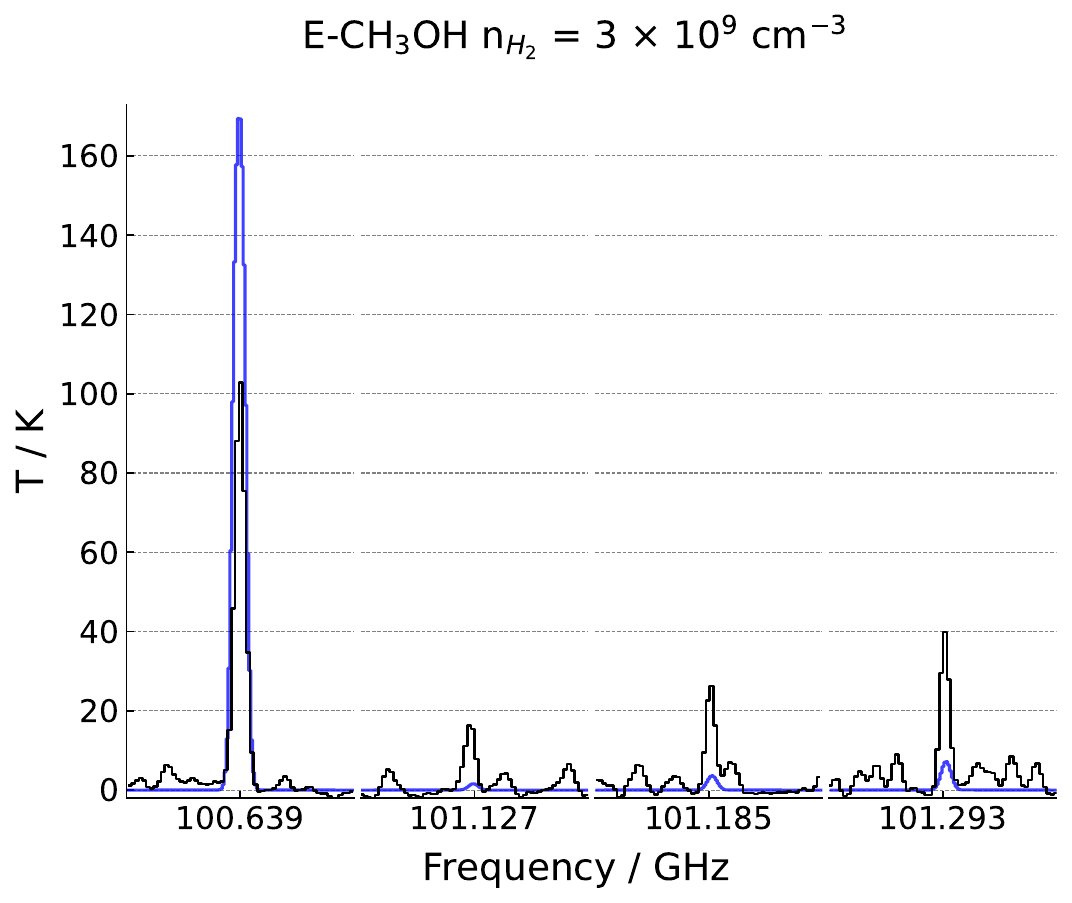}%
\hfill
\includegraphics[width=0.48\textwidth]{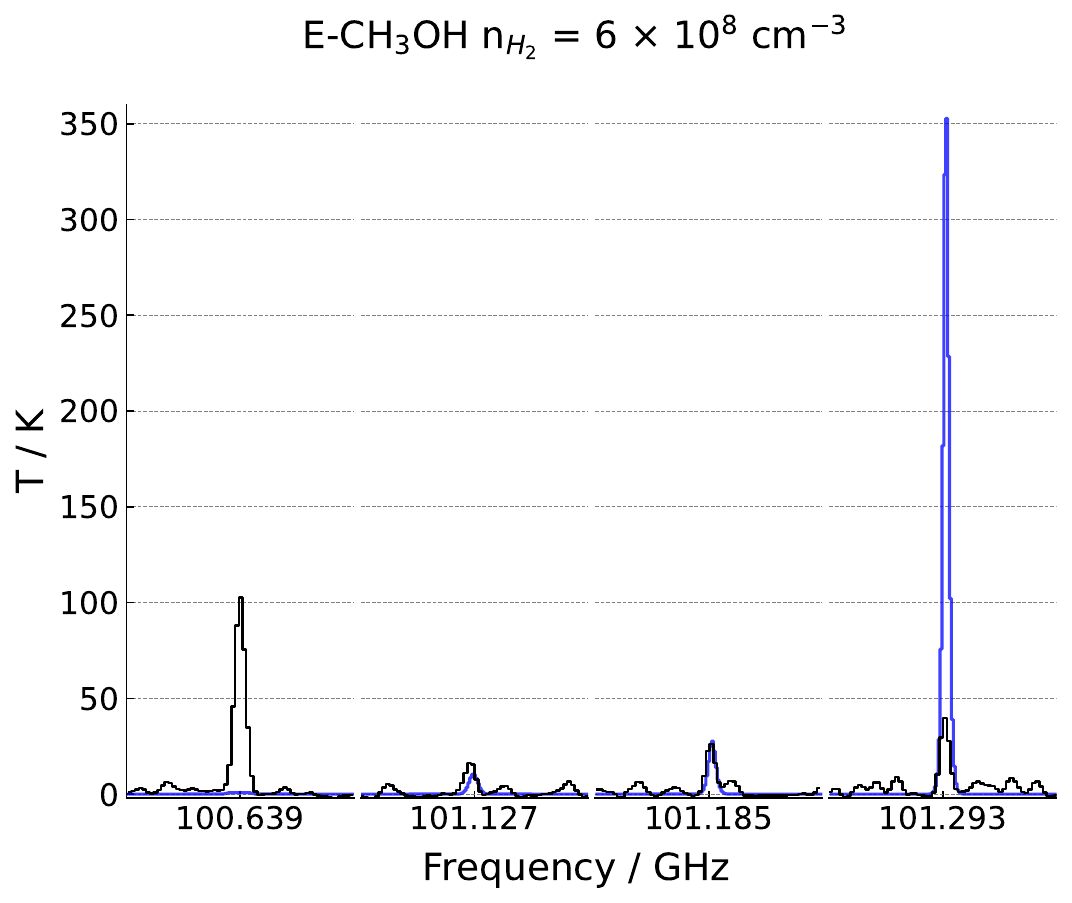}%
\caption{%
RADEX-calculated methanol line intensities are shown in blue, while the black spectra display the observed lines toward I16272-4837c1 (already divided by the beam filling factor).Both models adopt a column density of $2.7\times10^{18}~\mathrm{cm^{-2}}$ and a temperature of $230$~K. The left panel uses a hydrogen density of $3\times10^{9}~\mathrm{cm^{-3}}$, and the right panel $6\times10^{8}~\mathrm{cm^{-3}}$. 
}
\label{fig:radex}
\end{figure*}


\begin{figure}[htbp]
\centering
\includegraphics[width=0.43\textwidth]{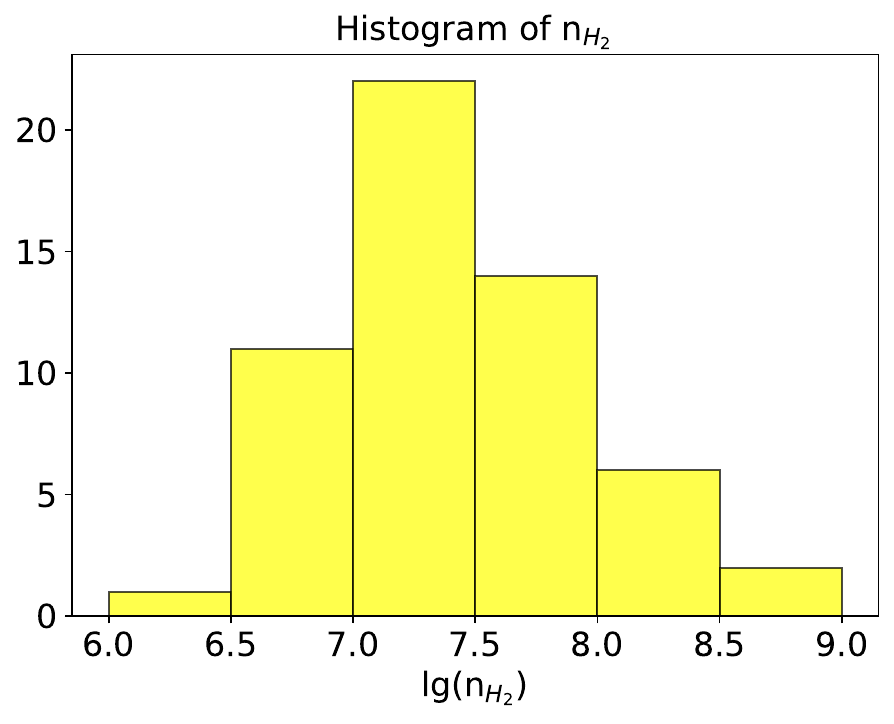}
\caption{Histogram of hydrogen density in hot core candidates without \ha.}
\label{Fig 5}
\end{figure}

\subsection{Molecular richness of hot molecular cores}
\label{sec 4.2}

\begin{figure*}[ht!]
\centering
\includegraphics[width=0.48\textwidth]{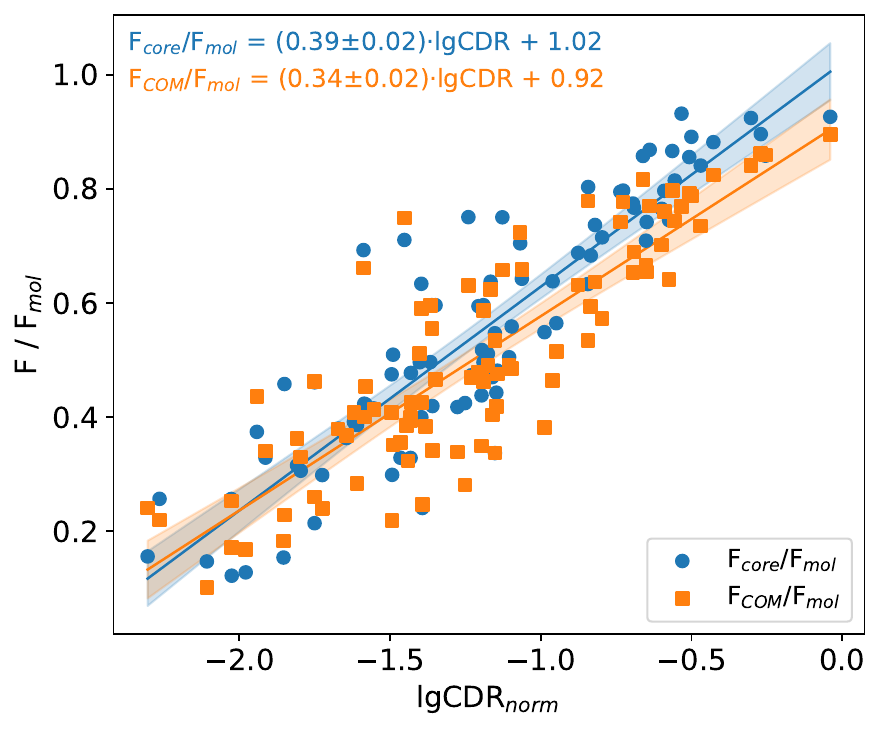}%
\hfill
\includegraphics[width=0.48\textwidth]{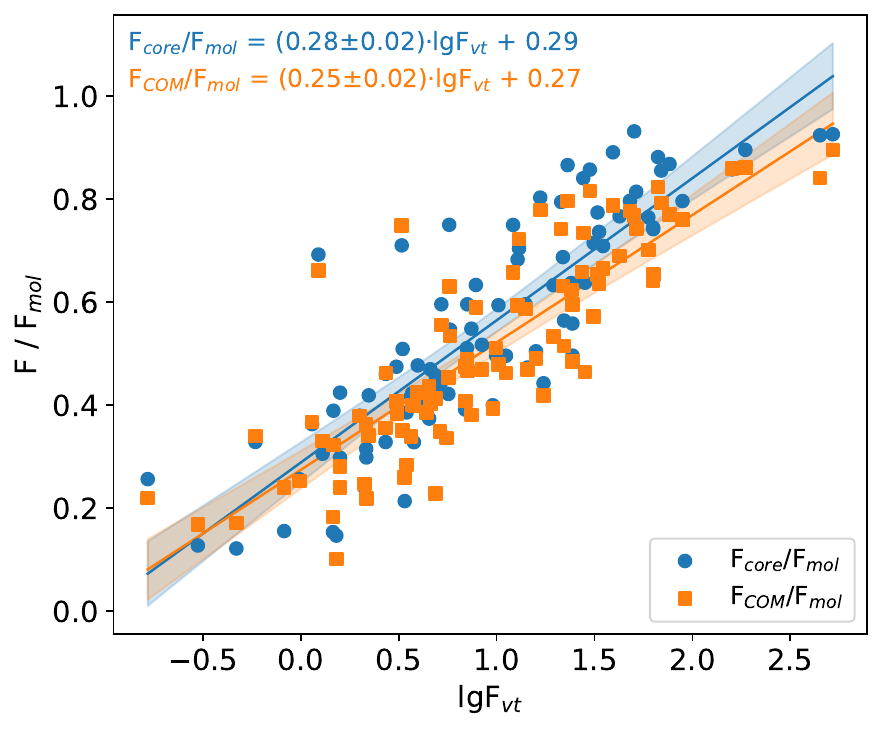}%
\caption{%
Chemical richness diagnostics for the 94 hot core candidates.
Top: Fractional contributions of hot-core molecules ($F_{core}/F_{mol}$) and complex organic molecules ($F_{COM}/F_{mol}$) as a function of the normalized Channel Detection Ratio (CDR${norm}$).
Bottom: Same fractions plotted against the integrated intensity of vibrationally excited methanol, $F{vt=1}$.
Both diagnostics show consistent positive correlations, with nearly identical trends for core-tracing and COM emission, indicating that molecular-line richness in hot cores is dominated by COMs and closely linked to warm, compact gas traced by vibrationally excited methanol.
}
\label{fig:ratio vs cdr}
\end{figure*}

Building on the quantitative measures introduced in Section~\ref{sec 3.3}, we further investigate how molecular line richness varies with the physical and chemical properties of hot cores. We calculate the total integrated intensities of: all molecular species ($F_{mol}$), molecules only originating in the hot core emission regions (hot core molecules; $F_{core}$), and all COMs ($F_{COM}$). We then calculate the ratios of $F_{core}/F_{mol}$ and $F_{COM}/F_{mol}$. Fig.~\ref{fig:ratio vs cdr} presents how these ratios change with two complementary parameters: the normalized CDR, CDR$_{norm}$, which measures the fraction of velocity channels with significant emission after correcting for line-width effects (see Appendix~D), and the integrated intensity of vibrationally excited methanol, $F_{vt=1}$, which traces the warmest and most highly excited gas.

Both the hot core molecular fraction ($F_{\rm core}/F_{\rm mol}$) and the COM fraction ($F_{\rm COM}/F_{\rm mol}$) exhibit clear positive correlations against $\rm CDR_{\rm norm}$ (upper panel). Linear fits yield $F_{\rm core}/F_{\rm mol}=(0.39\pm0.02)\log{\rm CDR}_{\rm norm}+1.02$ and $F_{\rm COM}/F_{\rm mol}=(0.34\pm0.02)\log{\rm CDR}_{\rm norm}+0.92$
with comparable dispersions. The close similarity in slope and scatter indicates that increasing spectral detectability is accompanied by a proportional rise in both hot core molecular emission and COM emission. The lack of a significant distinction between the two trends suggests that, within hot cores, chemically rich spectra are overwhelmingly dominated by COM transitions.

A consistent picture emerges in the correlation with $\log F_{vt=1}$ (lower panel of Fig.~\ref{fig:ratio vs cdr}). Both $F_{core}/F_{mol}$ and $F_{COM}/F_{mol}$ increase monotonically with $F_{vt=1}$, with slopes of 0.28 and 0.25, respectively. As vibrationally excited methanol selectively traces compact, warm, and dense gas close to the heating sources, this correlation demonstrates that the enhancement of chemical richness is closely linked to the excitation conditions in the inner regions of hot cores. Cores with stronger \vt\ methanol emission systematically show higher relative contributions from COMs and hot core molecules.

The near-indistinguishable behavior of $F_{core}/F_{mol}$ and $F_{COM}/F_{mol}$ in both diagnostics is physically expected. In hot-core environments, most species classified as core tracers belong to the COM family, and the molecular emission budget is therefore largely governed by COM transitions. As a result, metrics that quantify either spectral detectability (CDR$_{norm}$) or high-excitation gas ($F_{vt=1}$) naturally capture the same underlying chemical structure: a warm, compact region enriched in molecules.

Taken together, these results indicate that molecular-line richness in hot cores is not driven by a small number of exceptionally bright species, but instead reflects a global enhancement of chemically complex emission associated with warm, dense gas. Both CDR$_{norm}$ and vibrationally excited methanol provide robust, physically motivated proxies for this richness, enabling homogeneous comparisons across large samples and reinforcing the view that hot cores represent a chemically distinct evolutionary stage.


\section{Conclusions}
\label{conclution}

Based on the data obtained from the ATOMS project, we conducted a systematic survey of methanol (\meoh) emission lines within dense cores. The key findings of our study are summarized as follows:

1. We identified a total of 20 methanol transitions in the ATOMS 3 mm band, and identified 94 hot core candidates showing compact and strong \vteme~emission.

2. Among the 94 cores, 87 show the presence of COMs other than methanol. The “richness” of molecules is found to be highly correlated with the emission intensity of \vteme, suggesting that \vt~can serve as a reliable tracer for the richness of molecular lines as well as hot cores. While the CDR$_{core}$ can serve as a convenient and relatively reliable parameter for measuring the richness of molecules.

3. In these 94 cores, the rotational temperatures of \ame, \eme, and \vteme~showed distinct differences:
   \begin{itemize}
       \item The highest average rotational temperature of 194 $\pm$ 33 K was found for \vteme.
       \item \ame~exhibited an average rotational temperature of 178 $\pm$ 33 K.
       \item \eme~had a significantly lower average rotational temperature of 75 $\pm$ 21 K.
   \end{itemize}
   This temperature discrepancy arises from the enhanced \ejj~(J=3–7) E-type lines, which are stronger than predicted and concentrated near hot core centers. Their behavior biases the fitting results, highlighting the necessity of calculating E-type transitions separately from A-type and vibrationally excited lines. Our non-LTE calculations also show that densities exceeding the typical values found in hot cores are required for these transitions to become fully thermalized, implying that not all lines detected in hot cores can be treated under the simple LTE assumption.

These findings enhance our understanding of astrochemical processes in the early stages of massive star formation, offering new perspectives on the role of methanol and other COMs in star-forming regions. Future high-resolution observations and laboratory experiments will be crucial to further explore the excitation mechanisms and chemical evolution of methanol under varying interstellar conditions.

\noindent


\begin{acknowledgments}
Tie Liu acknowledges the supports by the National Key R\&D
Program of China (No. 2022YFA1603101), National Natural Science Foundation of China (NSFC) through grants No.12073061 and No.12122307, the PIFI program of Chinese Academy of Sciences through grant No. 2025PG0009, and the Tianchi Talent Program of Xinjiang
Uygur Autonomous Region.\\

S.-L. Qin is supported by National Natural Science Foundation of China (NSFC) through grant No.12033005.\\

Y.P. Peng acknowledges support from NSFC through grant No. 12303028.\\

SRD acknowledges support from the Fondecyt Postdoctoral fellowship (project code 3220162) and ANID BASAL project FB210003.\\

PS was partially supported by a Grant-in-Aid for Scientific Research (KAKENHI Number JP22H01271 and JP23H01221) of JSPS.\\

MJ acknowledges the support of the Research Council of Finland Grant No. 348342.\\

G.G. and L. B. gratefully acknowledge support by the ANID BASAL project FB210003.\\

C.W.L is supported by the Basic Science Research Program through the NRF funded by the Ministry of Education, Science and Technology (grant No. NRF-2019R1A2C1010851) and by the Korea Astronomy and Space Science Institute grant funded by the Korea government (MSIT; project No. 2025-1-841-02).\\

\end{acknowledgments}




\appendix
\section{Core positions and parameter of \meoh~lines} \label{app:A}

\begin{longrotatetable}
\begin{deluxetable*}{lccccccccccccccccl} 
\setlength{\tabcolsep}{0.5pt}
\tablecaption{Core positions and parameter of \meoh~lines \label{Table 2}}
\tablewidth{0pt} 
\tablehead{
\colhead{Source} & \colhead{RA} & \colhead{Dec} & \colhead{$\theta_{source}$} & 
\multicolumn{4}{c}{\ame} & \multicolumn{4}{c}{\eme} & \multicolumn{4}{c}{\vteme} & \colhead{H$_{40}\alpha$(I/O)} \\
\cline{5-17}
\colhead{IRAS} & \colhead{h.m.s} & \colhead{° ' "} & \colhead{"} & 
\colhead{$T_{rot}$} & \colhead{$N_{tot}$} & \colhead{$V_{width}$} & \colhead{$V_{off}$} &
\colhead{$T_{rot}$} & \colhead{$N_{tot}$} & \colhead{$V_{width}$} & \colhead{$V_{off}$} &
\colhead{$T_{rot}$} & \colhead{$N_{tot}$} & \colhead{$V_{width}$} & \colhead{$V_{off}$} & \colhead{or lg(\nh)} \\
\colhead{} & \colhead{} & \colhead{} & \colhead{} &
\colhead{K} & \colhead{10$^{16}$cm$^{-2}$} & \colhead{\kms} & \colhead{\kms} &
\colhead{K} & \colhead{10$^{16}$cm$^{-2}$} & \colhead{\kms} & \colhead{\kms} &
\colhead{K} & \colhead{10$^{16}$cm$^{-2}$} & \colhead{\kms} & \colhead{\kms} & \colhead{cm$^{-3}$}
}
\startdata
08303$-$4303 & 08:32:08.68 & $-$43:13:45.78 & 1.53 & 110 & 23 $\pm$ 1.6 & 5.7 & 12.8 & 75 $\pm$ 2 & 24 $\pm$ 3 & 5.3 & 12 & 184 $\pm$ 10 & 8 $\pm$ 0.5 & 5.5 & 10.5 & 7.4\\
08470$-$4243 & 08:48:47.79 & $-$42:54:27.90 & 1.45 & 204 $\pm$ 8 & 85 $\pm$ 6 & 6 & 13.2 & 58 $\pm$ 2 & 170 $\pm$ 10 & 5.5 & 13.6 & 194 $\pm$ 0.04 & 70 $\pm$ 0.8 & 6.3 & 13.9 & 7.7\\
09018$-$4816 & 09:03:33.46 & $-$48:28:01.69 & 1.64 & 148 $\pm$ 6 & 20 $\pm$ 2 & 6.3 & 10.4 & 66 $\pm$ 3 & 30 $\pm$ 3 & 5.7 & 10.2 & 157 $\pm$ 5 & 17.5 $\pm$ 1.5 & 6.3 & 10.3 & 7.6\\
10365$-$5803$^{N,x}$ & 10:38:32.15 & $-$58:19:08.30 & 2.33 & 146 $\pm$ 4 & 5.9 $\pm$ 0.5 & 4.7 & -19.5 & 69 $\pm$ 3 & 8.2 $\pm$ 0.5 & 4.3 & -19.8 & 183 $\pm$ 4 & 3.9 $\pm$ 0.4 & 4.3 & -19.6 & 7.1\\
11298$-$6155 & 11:32:05.59 & $-$62:12:25.62 & 2.09 & 153 $\pm$ 5 & 14 $\pm$ 2 & 6.3 & 32.8 & 63 $\pm$ 3 & 20 $\pm$ 3 & 6.1 & 33.2 & 199 $\pm$ 4 & 11.5 $\pm$ 1.5 & 7 & 33 & 6.6\\
11332$-6258^{N,x}$ & 11:35:32.27 & $-$63:14:42.82 & 1.08$^a$ & 155 $\pm$ 5 & 15 $\pm$ 3 & 6.9 & -13.6 & 67 & 17 $\pm$ 5 & 5.2 & -12.6 & 177 $\pm$ 6 & 8.5 $\pm$ 1.5 & 7 & -10.6 & \\
12320$-6122^{N,x}$ & 12:34:53.26 & $-$61:39:40.30 & 1.66$^a$ &  &  &  &  & 100 & 9 $\pm$ 1 & 6 & -38.5 & 196 $\pm$ 7 & 4.5 $\pm$ 0.5 & 6 & -38.5 & I\\
12326$-$6245 & 12:35:35.10 & $-$63:02:30.53 & 2.28$^a$ &  &  &  &  & 120 & 18 $\pm$ 2 & 4 & -41.5 & 215 $\pm$ 9 & 9 $\pm$ 1 & 6 & -40.5 & I\\
13079$-$6218c1 & 13:11:13.75 & $-$62:34:41.56 & 2 &  &  &  &  & 67 $\pm$ 2 & 140 $\pm$ 15 & 6.4 & -40.2 & 192 $\pm$ 7 & 85 $\pm$ 5 & 8 & -40.2 & 7.6\\
13079$-6218c2^N$ & 13:11:10.43 & $-$62:34:38.91 & 1.86 &  &  &  &  & 63 $\pm$ 3 & 25 $\pm$ 4 & 5.2 & -41.3 & 196 $\pm$ 5 & 12.5 $\pm$ 2 & 6 & -41.3 & 6.9\\
13134$-6242$ & 13:16:43.20 & $-$62:58:32.50 & 2 &  &  &  &  & 58 $\pm$ 3 & 190 $\pm$ 20 & 6.8 & -33.9 & 206 $\pm$ 6 & 95 $\pm$ 7.5 & 7 & -34.2 & 7.4\\
13140$-6226$ & 13:17:15.50 & $-$62:42:23.60 & 3.56 &  &  &  &  & 54 $\pm$ 3 & 12.7 $\pm$ 2.3 & 5.7 & -33.4 & 165 $\pm$ 6 & 4.7 $\pm$ 0.7 & 5.9 & -33.3 & 6.5\\
13471$-6120$ & 13:50:41.76 & $-$61:35:10.81 & 1.29 &  &  &  &  & 68 & 74 $\pm$ 8 & 3.6 & -59.8 & 172 $\pm$ 7 & 28 $\pm$ 4 & 4 & -59.8 & I\\
13484$-6100$ & 13:51:58.32 & $-$61:15:41.50 & 2.29 &  &  &  &  & 73 $\pm$ 4 & 9.2 $\pm$ 1 & 4 & -57 & 210 $\pm$ 8 & 4.9 $\pm$ 0.5 & 4.3 & -57 & 6.7\\
14164$-6028^{N,x}$ & 14:20:08.61 & $-$60:42:01.00 & 3.14 &  &  &  &  & 75 $\pm$ 5 & 8 $\pm$ 1.6 & 8.8 & -46.5 & 235 $\pm$ 5 & 4.7 $\pm$ 0.6 & 8 & -46.5 & 6.5\\
14212$-6131^N$ & 14:25:01.58 & $-$61:44:57.83 & 1.74 &  &  &  &  & 71 $\pm$ 3 & 21 $\pm$ 4 & 9 & -52 & 215 $\pm$ 9 & 16.5 $\pm$ 3.5 & 8 & -51.5 & 7.0\\
14498$-5856$ & 14:53:42.68 & $-$59:08:52.89 & 2.2 & 152 $\pm$ 9 & 28 $\pm$ 3 & 5.9 & -50.7 & 72 $\pm$ 7 & 45 $\pm$ 9 & 5.9 & -50.8 & 144 $\pm$ 7 & 29 $\pm$ 3 & 5.9 & -50.6 & 7.4\\
15254$-5621$ & 15:29:19.39 & $-$56:31:22.34 & 1.84$^a$ & 167 $\pm$ 6 & 42 $\pm$ 4 & 4.2 & -67.4 & 71 $\pm$ 7 & 95 $\pm$ 14 & 3.8 & -68.1 & 144 $\pm$ 10 & 31.5 $\pm$ 4 & 4 & -67.8 & I\\
15290$-5546^{N,x}$ & 15:32:52.86 & $-$55:56:06.75 & 3.13 & 142 $\pm$ 7 & 10 $\pm$ 2 & 8.3 & -88.5 & 79 $\pm$ 11 & 9 $\pm$ 3 & 7.9 & -89.2 & 216 $\pm$ 14 & 5 $\pm$ 0.5 & 6.5 & -89.3 & 6.7\\
15394$-5358^N$ & 15:43:16.66 & $-$54:07:14.49 & 3.25$^b$ & 194 $\pm$ 10 & 56 $\pm$ 5 & 6.5 & -39.8 & 53 $\pm$ 3 & 120 $\pm$ 23 & 6.8 & -39.9 & 189 $\pm$ 10 & 46.5 $\pm$ 3.5 & 7 & -39.9 & 7.5\\
15411$-5352^N$ & 15:44:59.58 & $-$54:02:22.63 & 1.75$^a$ & 200 $\pm$ 17 & 8.8 $\pm$ 1 & 6 & -42.5 & 55 & 14 $\pm$ 3 & 6.3 & -42.8 & 220 $\pm$ 13 & 7 $\pm$ 1.5 & 6.3 & -43.5 & O\\
15437$-5343$ & 15:47:32.73 & $-$53:52:38.80 & 2.08 & 151 $\pm$ 16 & 35 $\pm$ 6 & 6.1 & -85.3 & 88 $\pm$ 11 & 42 $\pm$ 4 & 5.9 & -85.3 & 161 $\pm$ 14 & 27.5 $\pm$ 4.5 & 5.8 & -85.3 & 6.9\\
15520$-5234$ & 15:55:48.47 & $-$52:43:06.75 & 2.56$^a$ & 148 $\pm$ 11 & 20 $\pm$ 3 & 4.3 & -43.6 & 75 $\pm$ 4 & 36 $\pm$ 5 & 3.6 & -43.6 & 174 $\pm$ 13 & 16.5 $\pm$ 1.5 & 4 & -43.6 & O\\
15557$-5215^N$ & 15:59:40.71 & $-$52:23:27.99 & 1.97$^b$ & 178 $\pm$ 12 & 9.6 $\pm$ 2 & 6.4 & -68.3 & 43 $\pm$ 4 & 16 $\pm$ 6 & 5.5 & -67.9 & 234 $\pm$ 12 & 3.25 $\pm$ 1.5 & 6.3 & -69 & 7.1\\
16037$-5223^N$ & 16:07:38.19 & $-$52:31:01.65 & 2.1 & 165 $\pm$ 10 & 13 $\pm$ 2.5 & 6 & -80.4 & 76 $\pm$ 6 & 7.8 $\pm$ 0.8 & 4.4 & -81 & 204 $\pm$ 14 & 6.5 $\pm$ 1.5 & 4.8 & -81.4 & 6.8\\
16060$-5146c1^R$ & 16:09:52.44 & $-$51:54:55.89 & 1.32$^a$ & 202 $\pm$ 7 & 143 $\pm$ 16 & 6.3 & -86.8 & 66 $\pm$ 4 & 240 $\pm$ 20 & 6 & -86.4 & 197 $\pm$ 8 & 130 $\pm$ 15 & 6 & -86.8 & O\\
16060$-5146c2^N$ & 16:09:52.72 & $-$51:54:53.28 & 1.62$^a$ & 145 $\pm$ 5 & 50 $\pm$ 7 & 5.6 & -95.8 & 113 $\pm$ 8 & 42 $\pm$ 7 & 5.2 & -96 & 157 $\pm$ 12 & 44 $\pm$ 3.5 & 6 & -95.8 & I\\
16065$-5158$ & 16:10:19.93 & $-$52:06:07.40 & 2.15$^a$ & 166 $\pm$ 8 & 90 $\pm$ 8 & 7.9 & -61.5 & 90 $\pm$ 7 & 100 $\pm$ 9 & 7.7 & -61.5 & 148 $\pm$ 11 & 86 $\pm$ 8 & 7.8 & -61.5 & I\\
16071$-5142^R$ & 16:10:59.75 & $-$51:50:22.81 & 2.86 & 190 $\pm$ 7 & 110 $\pm$ 12 & 6 & -87 & 57 $\pm$ 4 & 190 $\pm$ 20 & 6 & -87 & 204 $\pm$ 24 & 85 $\pm$ 9 & 6 & -86.9 & 7.4\\
16076$-5134$ & 16:11:26.56 & $-$51:41:57.19 & 1.72 & 162 $\pm$ 13 & 25 $\pm$ 3 & 7.5 & -87 & 81 & 19 $\pm$ 5 & 6 & -86 & 294 $\pm$ 23 & 11.5 $\pm$ 2 & 4.6 & -85 & 7.0\\
16119$-5048^N$ & 16:15:45.40 & $-$50:55:53.90 & 2.16$^b$ & 185 $\pm$ 14 & 7.6 $\pm$ 1 & 5.9 & -48.2 & 50 & 8.4 $\pm$ 1.5 & 4 & -48.2 & 235 $\pm$ 14 & 3.2 $\pm$ 0.4 & 4.3 & -48.2 & \\
16164$-5046$ & 16:20:11.08 & $-$50:53:14.75 & 2$^a$ & 190 $\pm$ 12 & 60 $\pm$ 6 & 5.2 & -58.9 & 78 $\pm$ 5 & 80 $\pm$ 9 & 4.5 & -59.2 & 182 $\pm$ 12 & 50 $\pm$ 5 & 4.5 & -59.2 & I\\
16172$-5028$ & 16:21:02.97 & $-$50:35:12.60 & 2.03$^a$ & 185 $\pm$ 12 & 24 $\pm$ 3 & 5.4 & -54.6 & 93 & 26 $\pm$ 4 & 6 & -54.2 & 220 & 13 $\pm$ 3 & 6 & -56.6 & I\\
16272$-4837c1$ & 16:30:58.77 & $-$48:43:53.57 & 1.6 & 250 $\pm$ 5 & 200 $\pm$ 7 & 4.5 & -46.6 & 68 $\pm$ 11 & 550 $\pm$ 60 & 4.5 & -46.6 & 227 $\pm$ 8 & 170 $\pm$ 6 & 5 & -46.6 & 7.8\\
16272$-4837c2$ & 16:30:58.65 & $-$48:43:51.27 & 1.23$^a$ & 167 $\pm$ 12 & 31 $\pm$ 10 & 5.4 & -46.6 & 64 $\pm$ 7 & 30 $\pm$ 10 & 5.3 & -46.6 & 205 $\pm$ 21 & 17 $\pm$ 5.5 & 5.8 & -46.6 & \\
16272$-4837c3$ & 16:30:57.29 & $-$48:43:39.87 & 1.45 & 174 $\pm$ 13 & 28 $\pm$ 4 & 5 & -46.6 & 48 $\pm$ 4 & 51 $\pm$ 10 & 5 & -46.8 & 242 $\pm$ 15 & 16.5 $\pm$ 4.5 & 5.5 & -46.8 & 7.0\\
16318$-4724$ & 16:35:33.96 & $-$47:31:11.59 & 1.9 & 190 $\pm$ 4 & 150 $\pm$ 7 & 6 & -119.8 & 76 $\pm$ 7 & 225 $\pm$ 35 & 6.6 & -120.3 & 188 $\pm$ 7 & 155 $\pm$ 10 & 7.9 & -120 & 7.3\\
16344$-4658$ & 16:38:09.49 & $-$47:04:59.73 & 1.47 & 159 $\pm$ 9 & 32 $\pm$ 4 & 5.5 & -49 & 60 $\pm$ 5 & 44 $\pm$ 5 & 4.8 & -49.2 & 206 $\pm$ 16 & 24 $\pm$ 4 & 4.8 & -49.5 & 7.1\\
16348$-4654c1$ & 16:38:29.64 & $-$47:00:35.97 & 1.18$^a$ & 223 $\pm$ 5 & 360 $\pm$ 10 & 7.5 & -49.5 & 85 $\pm$ 3 & 420 $\pm$ 20 & 7 & -49.5 & 229 $\pm$ 7 & 260 $\pm$ 15 & 8 & -49.5 & I\\
16348$-4654c2^N$ & 16:38:29.13 & $-$47:00:43.48 & 1.41 & 154 $\pm$ 12 & 29 $\pm$ 4 & 6.3 & -45.5 & 48 $\pm$ 7 & 24 $\pm$ 5 & 5.5 & -45.9 & 239 $\pm$ 23 & 12 $\pm$ 3.5 & 6.5 & -46.5 & 6.7\\
16351$-4722$ & 16:38:50.50 & $-$47:28:00.68 & 1.24 & 145 $\pm$ 12 & 55 $\pm$ 6 & 5.4 & -39.7 & 71 $\pm$ 7 & 62 $\pm$ 10 & 5.5 & -39.7 & 123 $\pm$ 16 & 31.5 $\pm$ 5.5 & 5.4 & -39.7 & 8.7\\
16424$-4531^N$ & 16:46:06.00 & $-$45:36:43.63 & 1.16$^a$ & 157 $\pm$ 11 & 25 $\pm$ 5 & 5.4 & -33.8 & 71 $\pm$ 6 & 38 $\pm$ 8 & 5.5 & -33.6 & 170 $\pm$ 18 & 21 $\pm$ 3.5 & 7 & -33.6 & 7.4\\
16445$-4459^N$ & 16:48:05.15 & $-$45:05:08.26 & 1.04$^a$ & 173 $\pm$ 13 & 38 $\pm$ 7 & 7.4 & -124.3 & 72 $\pm$ 7 & 51 $\pm$ 11 & 6.5 & -125.3 & 189 $\pm$ 14 & 27.5 $\pm$ 5.5 & 6.4 & -124.6 & O\\
16458$-4512$ & 16:49:30.04 & $-$45:17:44.58 & 0.84$^a$ & 220 $\pm$ 17 & 20 $\pm$ 5 & 3.5 & -51 & 54 $\pm$ 7 & 34 $\pm$ 3 & 3.5 & -51.3 & 167 $\pm$ 10 & 15.5 $\pm$ 1.5 & 4.8 & -51.3 & I\\
16484$-4603$ & 16:52:04.66 & $-$46:08:33.85 & 1.12 & 149 $\pm$ 15 & 96 $\pm$ 7 & 5 & -35.1 & 87 $\pm$ 10 & 130 $\pm$ 20 & 5.4 & -35.3 & 170 $\pm$ 15 & 75 $\pm$ 10 & 6.1 & -35.5 & 8.2\\
16547$-4247$ & 16:58:17.18 & $-$42:52:07.57 & 2.1 & 162 $\pm$ 11 & 29 $\pm$ 3 & 7.4 & -33.2 & 83 $\pm$ 6 & 26 $\pm$ 3 & 5.5 & -33.8 & 182 $\pm$ 14 & 18 $\pm$ 2.5 & 4.9 & -34 & 7.9\\
17008$-4040$ & 17:04:22.91 & $-$40:44:22.91 & 1.15 & 183 $\pm$ 5 & 390 $\pm$ 14 & 6.6 & -18.2 & 139 $\pm$ 9 & 400 $\pm$ 30 & 5.3 & -18.2 & 197 $\pm$ 11 & 290 $\pm$ 15 & 6.6 & -18.4 & 8.3\\
17016$-4124c1$ & 17:05:10.90 & $-$41:29:06.95 & 1.18 & 208 $\pm$ 5 & 370 $\pm$ 15 & 6.3 & -27.4 & 108 $\pm$ 18 & 670 $\pm$ 130 & 5.3 & -27.4 & 216 $\pm$ 9 & 285 $\pm$ 15 & 6.7 & -27.4 & 8.0\\
17016$-4124c2$ & 17:05:11.20 & $-$41:29:07.05 & 1.26$^a$ & 173 $\pm$ 11 & 28 $\pm$ 3 & 4.5 & -28 & 100 $\pm$ 16 & 29 $\pm$ 2 & 4 & -28 & 210 $\pm$ 22 & 17 $\pm$ 1.5 & 4.5 & -28.4 & I\\
17143$-3700^N$ & 17:17:45.46 & $-$37:03:11.77 & 2.93$^a$ & 154 $\pm$ 13 & 13 $\pm$ 1 & 4.7 & -31.1 & 53 $\pm$ 4 & 23 $\pm$ 2 & 4.5 & -31 & 177 & 7.5 $\pm$ 1 & 4.3 & -31 & I\\
17158$-3901c1$ & 17:19:20.43 & $-$39:03:51.58 & 2.2 & 138 $\pm$ 8 & 20 $\pm$ 1 & 5.2 & -16.6 & 90 $\pm$ 11 & 17 $\pm$ 2 & 4.5 & -16.6 & 135 $\pm$ 12 & 13.5 $\pm$ 1 & 5.1 & -16.4 & 7.5\\
17158$-3901c2$ & 17:19:20.47 & $-$39:03:49.20 & 1.58$^a$ & 173 $\pm$ 15 & 12 $\pm$ 1 & 5 & -18.6 & 56 $\pm$ 4 & 18 $\pm$ 1 & 3.8 & -19 & 171 $\pm$ 15 & 9 $\pm$ 1 & 4.7 & -18.6 & \\
17160$-3707^N$ & 17:19:27.42 & $-$37:11:07.64 & 0.77$^a$ & 182 $\pm$ 15 & 23 $\pm$ 2 & 7.5 & -68.5 & 52 $\pm$ 4 & 67 $\pm$ 6 & 6 & -68 & 218 $\pm$ 20 & 30 $\pm$ 3 & 7.5 & -68.2 & O\\
17175$-3544c1^R$ & 17:20:53.42 & $-$35:46:57.70 & 2.05 & 325 $\pm$ 30 & 1140 $\pm$ 250 & 6.4 & -6.1 & 83 $\pm$ 17 & 2600 $\pm$ 700 & 5 & -6.4 & 314 $\pm$ 20 & 900 $\pm$ 150 & 6.8 & -6.1 & 8.7\\
17175$-3544c2^N$ & 17:20:53.16 & $-$35:46:59.28 & 1.73 & 235 $\pm$ 26 & 420 $\pm$ 80 & 4.7 & -8.6 & 72 $\pm$ 19 & 1400 $\pm$ 300 & 4 & -8.6 & 224 $\pm$ 14 & 370 $\pm$ 55 & 4.7 & -8.6 & 8.1\\
17220$-3609$ & 17:25:25.22 & $-$36:12:45.34 & 2.35$^a$ & 183 $\pm$ 6 & 113 $\pm$ 6 & 6.8 & -97 & 73 $\pm$ 3 & 180 $\pm$ 12 & 5.8 & -96.3 & 174 $\pm$ 12 & 95 $\pm$ 5 & 6.7 & -96.8 & I\\
17233$-3606$ & 17:26:42.46 & $-$36:09:17.85 & 3.58 & 205 $\pm$ 4 & 75 $\pm$ 2 & 5 & -1.4 & 66 $\pm$ 3 & 97 $\pm$ 5 & 5.4 & -1.7 & 140 $\pm$ 4 & 55 $\pm$ 3.5 & 5.3 & -1.4 & 8.3\\
17278$-3541c1^{N,x}$ & 17:31:13.89 & $-$35:44:08.40 & 1.21 & 155 $\pm$ 14 & 19 $\pm$ 2 & 6.9 & -2.5 & 90 $\pm$ 8 & 15 $\pm$ 3 & 6.2 & -3 & 187 $\pm$ 10 & 6.5 $\pm$ 0.4 & 5 & -4 & 7.8\\
17278$-3541c2^N$ & 17:31:15.30 & $-$35:44:47.63 & 2.39 & 214 $\pm$ 8 & 58 $\pm$ 7 & 7.2 & 1 & 55 $\pm$ 4 & 80 $\pm$ 13 & 7.2 & 1 & 250 $\pm$ 30 & 43.5 $\pm$ 11.5 & 7.5 & 0 & 8.1\\
17439$-2845^N$ & 17:47:09.12 & $-$28:46:16.25 & 0.94 & 169 $\pm$ 14 & 31 $\pm$ 5 & 5.7 & 18.7 & 100 & 24 $\pm$ 2 & 5.7 & 18.5 & 210 $\pm$ 19 & 21 $\pm$ 2 & 6.5 & 18.4 & \\
17441$-2822c1^R$ & 17:47:20.08 & $-$28:23:06.48 & 1.5$^a$ & 225 $\pm$ 20 & 140 $\pm$ 13 & 8 & 66.7 & 68 $\pm$ 2 & 200 $\pm$ 18 & 7 & 66 & 196 $\pm$ 12 & 125 $\pm$ 6 & 7.5 & 66.3 & O\\
17441$-2822c2^N$ & 17:47:20.22 & $-$28:23:08.29 & 1.28$^a$ & 121 $\pm$ 7 & 65 $\pm$ 4 & 7.5 & 68 & 85 $\pm$ 4 & 100 $\pm$ 7 & 7.5 & 68 & 157 $\pm$ 24 & 70 $\pm$ 14.5 & 7.3 & 68 & O\\
17441$-2822c3^N$ & 17:47:20.46 & $-$28:23:46.24 & 1.17$^a$ & 189 $\pm$ 6 & 140 $\pm$ 8 & 5.5 & 70 & 98 $\pm$ 6 & 200 $\pm$ 20 & 5 & 69 & 186 $\pm$ 8 & 150 $\pm$ 10 & 6.3 & 69 & I\\
17441$-2822c4^N$ & 17:47:20.06 & $-$28:22:41.51 & 0.7$^a$ & 181 $\pm$ 8 & 200 $\pm$ 17 & 5.5 & 59.6 & 93 $\pm$ 8 & 330 $\pm$ 45 & 4.5 & 59.2 & 172 $\pm$ 11 & 200 $\pm$ 20 & 5.3 & 59.2 & I\\
17589$-2312^N$ & 18:01:57.74 & $-$23:12:34.26 & 0.91 & 154 $\pm$ 12 & 21 $\pm$ 2 & 7.3 & 19.3 & 51 $\pm$ 4 & 46 $\pm$ 4 & 5.7 & 19 & 235 $\pm$ 23 & 18.5 $\pm$ 1.5 & 8.6 & 19 & 7.0\\
17599$-2148^N$ & 18:03:00.74 & $-$21:48:10.12 & 2.48 & 171 $\pm$ 15 & 11 $\pm$ 1.1 & 5.3 & 20.3 & 69 $\pm$ 6 & 13 $\pm$ 1.2 & 5.2 & 20.4 & 197 $\pm$ 18 & 9.5 $\pm$ 1 & 6.4 & 20 & 7.4\\
18032$-2032c1$ & 18:06:14.94 & $-$20:31:43.22 & 1.76$^a$ & 143 $\pm$ 10 & 15 $\pm$ 1.4 & 4.5 & 4.7 & 114 $\pm$ 10 & 14 $\pm$ 1.1 & 4.8 & 4.7 & 242 $\pm$ 21 & 7.5 $\pm$ 0.7 & 4.7 & 4.7 & I\\
18032$-2032c2$ & 18:06:14.88 & $-$20:31:39.59 & 1.49 & 192 $\pm$ 9 & 59 $\pm$ 3 & 5.6 & 5.4 & 80 $\pm$ 6 & 50 $\pm$ 4 & 5.6 & 5.4 & 179 $\pm$ 15 & 39.5 $\pm$ 3 & 5.6 & 5.4 & 7.3\\
18032$-2032c3$ & 18:06:14.80 & $-$20:31:37.26 & 1.5$^a$ & 125 $\pm$ 11 & 14.5 $\pm$ 1.2 & 6 & 5 & 75 $\pm$ 6 & 19 $\pm$ 1.7 & 5.5 & 4.8 & 135 $\pm$ 13 & 12.5 $\pm$ 1.1 & 6 & 5.4 & \\
18032$-2032c4$ & 18:06:14.66 & $-$20:31:31.57 & 0.92 & 174 $\pm$ 7 & 170 $\pm$ 7 & 5 & 2 & 91 $\pm$ 3 & 220 $\pm$ 12 & 4.6 & 2 & 208 $\pm$ 14 & 110 $\pm$ 8 & 5.2 & 2 & 7.8\\
18056$-1952$ & 18:08:38.23 & $-$19:51:50.34 & 2$^a$ & 235 $\pm$ 1 & 1150 $\pm$ 15 & 10.2 & 66.7 & 160 $\pm$ 6 & 2070 $\pm$ 40 & 9.2 & 66.1 & 210 $\pm$ 2 & 1050 $\pm$ 10 & 10.2 & 66.7 & I\\
18089$-1732$ & 18:11:51.45 & $-$17:31:28.96 & 1.59 & 194 $\pm$ 5 & 155 $\pm$ 7 & 5.4 & 33 & 72 $\pm$ 2 & 310 $\pm$ 20 & 4.5 & 33 & 176 $\pm$ 5 & 155 $\pm$ 4 & 5.5 & 33 & 7.7\\
18117$-1753$ & 18:14:39.51 & $-$17:52:00.08 & 1.23 & 142 $\pm$ 9 & 95 $\pm$ 7 & 4.5 & 38 & 98 $\pm$ 5 & 170 $\pm$ 13 & 4 & 38.7 & 146 $\pm$ 13 & 85 $\pm$ 7 & 4.6 & 38.6 & 7.7\\
18134$-1942^{N,x}$ & 18:16:22.12 & $-$19:41:27.06 & 1.57 & 136 $\pm$ 10 & 18 $\pm$ 1.5 & 6 & 10.5 & 79 $\pm$ 6 & 33 $\pm$ 2.6 & 7.6 & 10.5 & 180 $\pm$ 16 & 17.5 $\pm$ 1.6 & 7.6 & 10.5 & 7.8\\
18159$-1648c1$ & 18:18:54.66 & $-$16:47:50.28 & 1.35 & 252 $\pm$ 15 & 58 $\pm$ 5 & 4.5 & 22 & 42 $\pm$ 2 & 190 $\pm$ 15 & 5 & 22 & 233 $\pm$ 19 & 44.5 $\pm$ 4 & 5 & 22 & 7.7\\
18159$-1648c2$ & 18:18:54.34 & $-$16:47:49.97 & 1.29 & 197 $\pm$ 12 & 52 $\pm$ 5 & 5 & 23 & 62 $\pm$ 3 & 94 $\pm$ 7 & 5 & 23 & 169 $\pm$ 15 & 54 $\pm$ 4.5 & 5.6 & 23 & 7.7\\
18182$-1433^R$ & 18:21:09.12 & $-$14:31:48.54 & 1.95 & 169 $\pm$ 7 & 45 $\pm$ 5 & 5.7 & 60 & 58 $\pm$ 3 & 73 $\pm$ 8 & 5 & 59.5 & 183 $\pm$ 8 & 32 $\pm$ 3.5 & 6.2 & 60 & 7.4\\
18236$-1205$ & 18:26:25.79 & $-$12:03:53.08 & 1.37 & 216 $\pm$ 17 & 22 $\pm$ 2 & 5.5 & 26.5 & 43 $\pm$ 3 & 41 $\pm$ 4 & 5 & 26.5 & 240 $\pm$ 23 & 8.5 $\pm$ 0.8 & 5.7 & 26.5 & 7.2\\
18264$-1152c1^N$ & 18:29:14.37 & $-$11:50:22.38 & 1.5 & 217 $\pm$ 21 & 24 $\pm$ 2 & 5.5 & 44.5 & 49 $\pm$ 3 & 46 $\pm$ 5 & 5.2 & 45.5 & 244 $\pm$ 22 & 19.5 $\pm$ 1.6 & 6 & 45.3 & 7.3\\
18264$-1152c2^N$ & 18:29:14.43 & $-$11:50:24.52 & 1.68 & 191 $\pm$ 8 & 16 $\pm$ 1 & 5 & 44.5 & 40 $\pm$ 2 & 40 $\pm$ 4 & 5.2 & 43.8 & 210 $\pm$ 19 & 13 $\pm$ 1.2 & 7.2 & 45 & 7.3\\
18290$-0924$ & 18:31:44.13 & $-$09:22:12.25 & 0.86 & 171 $\pm$ 15 & 34 $\pm$ 3 & 6.5 & 83.8 & 64 $\pm$ 5 & 59 $\pm$ 5 & 7 & 83.8 & 170 $\pm$ 15 & 36 $\pm$ 3 & 6.5 & 83.8 & 7.1\\
18316$-0602$ & 18:34:20.91 & $-$05:59:42.00 & 1.61 & 174 $\pm$ 7 & 80 $\pm$ 6 & 6.5 & 41.5 & 79 $\pm$ 4 & 102 $\pm$ 9 & 6 & 41.2 & 183 $\pm$ 15 & 65 $\pm$ 5 & 6.8 & 41.7 & 7.5\\
18411$-0338$ & 18:43:46.23 & $-$03:35:29.77 & 0.95$^a$ & 178 $\pm$ 12 & 57 $\pm$ 5 & 6.8 & 102.8 & 74 $\pm$ 5 & 72 $\pm$ 6 & 7.7 & 102.8 & 205 $\pm$ 18 & 42 $\pm$ 3 & 6.8 & 102 & 7.4\\
18434$-0242^N$ & 18:46:03.75 & $-$02:39:22.36 & 1.07 & 178 $\pm$ 4 & 370 $\pm$ 10 & 6.7 & 97.8 & 105 $\pm$ 2 & 540 $\pm$ 16 & 5.1 & 97.5 & 188 $\pm$ 4 & 270 $\pm$ 6 & 6.5 & 97.5 & \\
18461$-0113^N$ & 18:48:41.94 & $-$01:10:02.53 & 2.13 & 180 $\pm$ 14 & 12 $\pm$ 1 & 4.5 & 96 & 52 $\pm$ 4 & 19 $\pm$ 1.5 & 5.5 & 96 & 185 $\pm$ 16 & 6.5 $\pm$ 0.6 & 5.5 & 96 & \\
18469$-0132$ & 18:49:33.05 & $-$01:29:03.34 & 0.76 & 149 $\pm$ 10 & 99 $\pm$ 8 & 5.5 & 87.2 & 76 $\pm$ 4 & 180 $\pm$ 16 & 5.2 & 87 & 145 $\pm$ 12 & 100 $\pm$ 9 & 5.5 & 87 & I\\
18479$-0005^N$ & 18:50:30.73 & $-$00:01:59.31 & 1.37$^a$ & 154 $\pm$ 13 & 17 $\pm$ 1.6 & 7.4 & 11.6 & 100 $\pm$ 9 & 13 $\pm$ 1 & 6.4 & 11 & 159 $\pm$ 10 & 17 $\pm$ 1.4 & 7 & 10.3 & O\\
18507$+0110$ & 18:53:18.56 & $+$01:14:58.23 & 2.37$^a$ & 237 $\pm$ 2 & 350 $\pm$ 7 & 6.4 & 57.2 & 95 $\pm$ 11 & 780 $\pm$ 90 & 5.9 & 57.2 & 203 $\pm$ 3 & 310 $\pm$ 5 & 6.3 & 57.2 & I\\
18507$+0121$ & 18:53:18.01 & $+$01:25:25.56 & 1.47 & 201 $\pm$ 4 & 270 $\pm$ 8 & 7 & 58 & 85 $\pm$ 9 & 530 $\pm$ 50 & 6.3 & 58 & 190 $\pm$ 6 & 250 $\pm$ 8 & 7.3 & 58 & 8.0\\
18517$+0437$ & 18:54:14.24 & $+$04:41:40.65 & 1.56 & 173 $\pm$ 8 & 72 $\pm$ 5 & 6 & 44.5 & 82 $\pm$ 5 & 85 $\pm$ 7 & 6 & 44.5 & 185 $\pm$ 14 & 70 $\pm$ 6 & 6.8 & 44.5 & 7.3\\
19078$+0901c1$ & 19:10:13.16 & $+$09:06:14.17 & 1.46$^a$ & 168 & 6 & 5 & 19 & 68 & 19 & 7.2 & 16.6 & 140 & 4.25 & 5 & 19 & I\\
19078$+0901c2$ & 19:10:14.14 & $+$09:06:24.68 & 1.51$^a$ & 166 $\pm$ 10 & 64 $\pm$ 6 & 7.5 & 14 & 71 $\pm$ 4 & 100 $\pm$ 9 & 7.5 & 14 & 185 $\pm$ 12 & 70 $\pm$ 5.5 & 8 & 13.7 & \\
19078$+0901c3^N$ & 19:10:12.72 & $+$09:06:11.20 & 2.42$^a$ & 152 & 12 & 5 & 14 & 72 & 12 & 7.2 & 14.4 & 240 & 6 & 5 & 14.4 & \\
19095$+0930$ & 19:11:54.00 & $+$09:35:50.52 & 1.54$^a$ & 168 $\pm$ 10 & 29 $\pm$ 2 & 3.6 & 41.5 & 70 $\pm$ 4 & 43 $\pm$ 4 & 3.3 & 41.2 & 163 $\pm$ 14 & 25 $\pm$ 2.5 & 3.6 & 41 & I\\
\enddata
\end{deluxetable*}
\begin{flushleft}
\tablecomments{In the "Source" column, the superscript "N" denotes newly identified hot core candidates confirmed through the \vteme~line, while "R" indicates hot cores with positional corrections exceeding 1 beam. Core marked with “x” indicate sources with no COM lines above 3$\sigma$, which are therefore not classified as hot cores in this work. In the $\theta_{source}$ column, the superscript "a" means that the source size is determined by the line image of \eme~at 100639 MHz (the 5$^{th}$ subplot in Fig.~\ref{Fig 1. lineimages}), and "b" indicates that the source size and the data used for calculating the parameters to its right are from the ACA-12-m combined data. In the parameter calculation section, the error is not given for those with too few spectral lines to calculate, and estimated values are used instead. The rightmost column shows the detection of \ha~in the spectra. "I" indicates that the center of the methanol line image is within 1 beam of the \ha~emission center, while "O" indicates that the methanol line image is more than 1 beam away from the \ha~emission center or that the \ha~emission is diffuse. The numbers represent the logarithm of the hydrogen density (in units of cm$^{-3}$), which is calculated using formulas (B2) and $n_{\text{H}_2} = \frac{N_{\text{H}_2}}{2\theta_{\text{source}}}$ in Appendix B of \cite{2021MNRAS.505.2801L}. The blank space indicates that this core lacks \ha~emission, but the continuum is too irregular for Gaussian fitting, and thus \nh~cannot be calculated. The main reasons include: the core is too close to another hot core/H II region. Some of these candidates were not selected in \cite{2021MNRAS.505.2801L} and \cite{2022MNRAS.511.3463Q}.}
\end{flushleft}
\end{longrotatetable}

\begin{longrotatetable}
\begin{longtable}{lccccccccccccccc}
\caption{Detections of \meoh~lines} \label{Table 3} \\
\hline\hline
Core/Freq(MHz) & 97678 & 97856 & 99602 & 100585 & 99001 & 99374 & 99772 & 98267 & 101097 & 101101 & 101126 & 101185 & 101293 & counts$^{a}$ & $CDR_{norm}$ \\
\hline
\endfirsthead
\caption[]{Detections of \meoh~lines (continued)} \\
\hline\hline
Core/Freq(MHz) & 97678 & 97856 & 99602 & 100585 & 99001 & 99374 & 99772 & 98267 & 101097 & 101101 & 101126 & 101185 & 101293 & counts$^{a}$ & $CDR_{norm}$ \\
\hline
\endhead
\hline \multicolumn{16}{r}{\textit{Continued on next page}} \\
\endfoot
\hline
\endlastfoot
08303$  -4303		$	&		&		&		&		&		&		&	1	&		&		&		&		&		&	1	&	6	&	0.0414 	\\
08470$  -4243		$	&	1	&	1	&	1	&		&	1	&	1	&	1	&		&		&	1	&	1	&	1	&	1	&	17	&	0.0866 	\\
09018$  -4816		$	&	1	&		&		&		&		&		&	1	&		&		&		&		&		&	1	&	8	&	0.0263 	\\
10365$  -5803^x		$	&	1	&		&		&		&		&		&	1	&		&		&		&	1	&	1	&	1	&	10	&	0.0094 	\\
11298$  -6155		$	&	1	&		&		&		&		&		&	1	&		&		&		&		&	1	&	1	&	9	&	0.0178 	\\
11332$  -6258^x		$	&	1	&		&		&		&		&		&	1	&		&		&		&		&		&		&	7	&	0.0050 	\\
12320$  -6122^x		$	&	F	&		&		&		&		&		&	1	&		&		&		&		&		&		&	5	&	0.0054 	\\
12326$  -6245		$	&	F	&		&		&		&		&		&	1	&		&		&		&		&		&	1	&	6	&	0.0363 	\\
13079$  -6218	c1	$	&	F	&		&		&	1	&	1	&	1	&	1	&		&		&	1	&	1	&	1	&	1	&	13	&	0.2021 	\\
13079$	-6218	c2	$	&	F	&		&		&		&		&		&	1	&		&		&		&		&	1	&	1	&	7	&	0.0324 	\\
13134$	-6242		$	&	F	&		&	1	&		&	1	&	1	&	1	&		&		&	1	&	1	&	1	&	1	&	13	&	0.1513 	\\
13140$	-6226		$	&	F	&		&		&		&		&		&	1	&		&		&		&		&	1	&	1	&	7	&	0.0189 	\\
13471$	-6120		$	&	F	&		&		&		&		&		&	1	&		&		&		&	1	&	1	&	1	&	8	&	0.0575 	\\
13484$	-6100		$	&	F	&		&		&		&		&		&	1	&		&		&		&		&		&	1	&	6	&	0.0561 	\\
14164$	-6028^x		$	&	F	&		&		&		&		&		&	1	&		&		&		&		&		&	1	&	6	&	0.0078 	\\
14212$	-6131		$	&	F	&		&		&		&		&		&	1	&		&		&		&		&	1	&	1	&	7	&	0.0212 	\\
14498$	-5856		$	&	1	&		&		&		&		&		&	1	&		&		&		&	1	&	1	&	1	&	10	&	0.0369 	\\
15254$	-5621		$	&	1	&		&		&		&	1	&		&	1	&		&		&		&	1	&	1	&	1	&	11	&	0.0855 	\\
15290$	-5546^x		$	&	1	&		&		&		&		&		&	1	&		&		&		&		&		&		&	7	&	0.0321 	\\
15394$	-5358		$	&	1	&	1	&		&	1	&	1	&	1	&	1	&		&		&	1	&	1	&	1	&	1	&	17	&	0.0681 	\\
15411$	-5352		$	&	1	&		&		&		&		&		&	1	&		&		&		&		&		&		&	7	&	0.0160 	\\
15437$	-5343		$	&	1	&		&		&		&		&	1	&	1	&		&		&		&		&	1	&	1	&	11	&	0.0397 	\\
15520$	-5234		$	&	1	&		&		&		&		&		&	1	&		&		&		&	1	&	1	&	1	&	10	&	0.0717 	\\
15557$	-5215		$	&	1	&		&		&		&		&		&	1	&		&		&		&		&	1	&	1	&	9	&	0.0105 	\\
16037$	-5223		$	&	1	&		&		&		&		&		&	1	&		&		&		&		&		&	1	&	8	&	0.0302 	\\
16060$	-5146	c1	$	&	1	&	1	&	1	&	1	&	1	&	1	&	1	&	1	&		&	1	&	1	&	1	&	1	&	19	&	0.2234 	\\
16060$	-5146	c2	$	&	1	&		&		&		&		&		&	1	&		&		&		&		&		&	1	&	8	&	0.1095 	\\
16065$	-5158		$	&	1	&		&		&		&	1	&		&	1	&		&		&		&	1	&	1	&	1	&	11	&	0.1598 	\\
16071$	-5142		$	&	1	&	1	&	1	&	1	&	1	&	1	&	1	&	1	&		&	1	&	1	&	1	&	1	&	19	&	0.2037 	\\
16076$	-5134		$	&	1	&		&		&		&		&		&	1	&		&		&		&		&		&	1	&	8	&	0.0636 	\\
16119$	-5048		$	&	1	&		&		&		&		&		&	1	&		&		&		&		&	1	&	1	&	9	&	0.0140 	\\
16164$	-5046		$	&	1	&	1	&	1	&		&	1	&	1	&	1	&		&		&		&	1	&	1	&	1	&	16	&	0.1836 	\\
16172$	-5028		$	&	1	&		&		&		&		&	1	&	1	&		&		&		&		&	1	&	1	&	11	&	0.0448 	\\
16272$	-4837	c1	$	&	1	&	1	&	1	&	1	&	1	&	1	&	1	&	1	&	1	&	1	&	1	&	1	&	1	&	20	&	0.3737 	\\
16272$	-4837	c2	$	&	1	&		&		&		&		&		&	1	&		&		&		&	1	&	1	&	1	&	10	&	0.0668 	\\
16272$	-4837	c3	$	&	1	&		&		&		&	1	&	1	&	1	&		&		&		&	1	&	1	&	1	&	13	&	0.0704 	\\
16318$	-4724		$	&	1	&	1	&	1	&	1	&	1	&	1	&	1	&		&		&	1	&	1	&	1	&	1	&	18	&	0.2245 	\\
16344$	-4658		$	&	1	&		&		&		&		&		&	1	&		&		&		&	1	&	1	&	1	&	10	&	0.0639 	\\
16348$	-4654	c1	$	&	1	&	1	&	1	&	1	&	1	&	1	&	1	&	1	&		&	1	&	1	&	1	&	1	&	19	&	0.2298 	\\
16348$	-4654	c2	$	&	1	&		&	1	&		&	1	&		&	1	&		&		&		&	1	&	1	&	1	&	12	&	0.0370 	\\
16351$	-4722		$	&	1	&		&	1	&		&		&		&	1	&		&		&		&		&	1	&	1	&	10	&	0.0704 	\\
16424$	-4531		$	&	1	&		&		&		&		&		&	1	&		&		&		&		&	1	&	1	&	9	&	0.0246 	\\
16445$	-4459		$	&	1	&		&		&		&		&		&	1	&		&		&		&		&	1	&	1	&	9	&	0.0280 	\\
16458$	-4512		$	&	1	&		&		&		&		&		&	1	&		&		&		&		&		&	1	&	8	&	0.0258 	\\
16484$	-4603		$	&	1	&		&	1	&		&	1	&	1	&	1	&		&		&		&		&	1	&	1	&	13	&	0.0589 	\\
16547$	-4247		$	&	1	&		&		&		&	1	&		&	1	&		&		&		&		&	1	&	1	&	10	&	0.1029 	\\
17008$	-4040		$	&	1	&		&	1	&		&	1	&	1	&	1	&		&		&	1	&	1	&	1	&	1	&	15	&	0.2518 	\\
17016$	-4124	c1	$	&	1	&	1	&	1	&		&	1	&	1	&	1	&		&		&	1	&	1	&	1	&	1	&	17	&	0.0353 	\\
17016$	-4124	c2	$	&	1	&		&		&		&	1	&		&	1	&		&		&		&		&		&	1	&	9	&	0.0437 	\\
17143$	-3700		$	&	1	&		&		&		&		&		&	1	&		&		&		&		&	1	&	1	&	9	&	0.0691 	\\
17158$	-3901	c1	$	&	1	&		&		&		&		&		&	1	&		&		&		&		&	1	&	1	&	9	&	0.0401 	\\
17158$	-3901	c2	$	&	1	&		&		&		&		&		&	1	&		&		&		&		&		&	1	&	8	&	0.0437 	\\
17160$	-3707		$	&	1	&		&		&		&		&		&	1	&		&		&		&		&	1	&	1	&	9	&	0.0178 	\\
17175$	-3544	c1	$	&	1	&	1	&	1	&	1	&	1	&	1	&	1	&	1	&	1	&	1	&	1	&	1	&	1	&	20	&	0.9122 	\\
17175$	-3544	c2	$	&	1	&	1	&	1	&	1	&	1	&	1	&	1	&	1	&	1	&	1	&	1	&	1	&	1	&	20	&	0.5561 	\\
17220$	-3609		$	&	1	&	1	&		&		&	1	&	1	&	1	&		&		&	1	&	1	&	1	&	1	&	16	&	0.1876 	\\
17233$	-3606		$	&	1	&	1	&		&		&		&	1	&	1	&		&		&	1	&	1	&	1	&	1	&	15	&	0.3392 	\\
17278$	-3541	c1^x	$	&	1	&		&		&		&		&		&	1	&		&		&		&		&		&		&	7	&	0.0122 	\\
17278$	-3541	c2	$	&	1	&		&		&		&	1	&	1	&	1	&		&		&		&		&	1	&	1	&	12	&	0.2182 	\\
17439$	-2845		$	&	1	&		&		&		&		&		&	1	&		&		&		&		&		&	1	&	8	&	0.0260 	\\
17441$	-2822	c1	$	&	1	&	1	&		&		&	1	&	1	&	1	&		&		&	1	&	1	&	1	&	1	&	16	&	0.2930 	\\
17441$	-2822	c2	$	&	1	&		&		&		&		&		&	1	&		&		&		&	1	&		&	1	&	9	&	0.0746 	\\
17441$	-2822	c3	$	&	1	&		&		&		&		&		&	1	&		&		&		&		&	1	&	1	&	9	&	0.3159 	\\
17441$	-2822	c4	$	&	1	&		&		&		&		&	1	&	1	&		&		&		&	1	&	1	&	1	&	12	&	0.2730 	\\
17589$	-2312		$	&	1	&		&		&		&		&		&	1	&		&		&		&		&	1	&	1	&	9	&	0.0156 	\\
17599$	-2148		$	&	1	&		&		&		&		&		&	1	&		&		&		&		&	1	&	1	&	9	&	0.0371 	\\
18032$	-2032	c1	$	&	1	&		&		&		&		&		&	1	&		&		&		&		&		&	1	&	8	&	0.0343 	\\
18032$	-2032	c2	$	&	1	&		&		&		&	1	&	1	&	1	&		&		&		&		&	1	&	1	&	12	&	0.1465 	\\
18032$	-2032	c3	$	&	1	&		&		&		&		&		&	1	&		&		&		&		&	1	&	1	&	9	&	0.0529 	\\
18032$	-2032	c4	$	&	1	&		&	1	&		&		&	1	&	1	&		&		&		&	1	&	1	&	1	&	13	&	0.1331 	\\
18056$	-1952		$	&	1	&	1	&	1	&	1	&	1	&	1	&	1	&		&		&		&	1	&	1	&	1	&	17	&	0.4980 	\\
18089$	-1732		$	&	1	&	1	&	1	&		&	1	&	1	&	1	&		&	1	&	1	&	1	&	1	&	1	&	18	&	0.2782 	\\
18117$	-1753		$	&	1	&		&	1	&		&	1	&		&	1	&		&		&	1	&	1	&	1	&	1	&	13	&	0.1433 	\\
18134$	-1942^x		$	&	1	&		&		&		&		&		&	1	&		&		&		&		&	1	&	1	&	9	&	0.0114 	\\
18159$	-1648	c1	$	&	1	&	1	&	1	&		&	1	&	1	&	1	&		&		&	1	&	1	&	1	&	1	&	17	&	0.1433 	\\
18159$	-1648	c2	$	&	1	&		&		&		&		&		&	1	&		&		&		&	1	&	1	&	1	&	10	&	0.0645 	\\
18182$	-1433		$	&	1	&		&		&		&	1	&	1	&	1	&		&		&		&		&	1	&	1	&	12	&	0.0785 	\\
18236$	-1205		$	&	1	&		&		&		&		&		&	1	&		&		&		&		&		&	1	&	8	&	0.0320 	\\
18264$	-1152	c1	$	&	1	&		&		&		&		&		&	1	&		&		&		&		&	1	&	1	&	9	&	0.0240 	\\
18264$	-1152	c2	$	&	1	&		&		&		&		&		&	1	&		&		&		&		&	1	&	1	&	9	&	0.0227 	\\
18290$	-0924		$	&	1	&		&		&		&		&		&	1	&		&		&		&		&		&	1	&	8	&	0.0436 	\\
18316$	-0602		$	&	1	&		&	1	&		&		&	1	&	1	&		&		&		&	1	&	1	&	1	&	13	&	0.0801 	\\
18411$	-0338		$	&	1	&		&		&		&	1	&	1	&	1	&		&		&		&		&	1	&	1	&	12	&	0.0622 	\\
18434$	-0242		$	&	1	&	1	&	1	&	1	&	1	&	1	&	1	&		&		&	1	&	1	&	1	&	1	&	18	&	0.2665 	\\
18461$	-0113		$	&	1	&		&		&		&		&		&	1	&		&		&		&		&	1	&	1	&	9	&	0.0213 	\\
18469$	-0132		$	&	1	&		&		&		&		&		&	1	&		&		&		&	1	&	1	&	1	&	10	&	0.0646 	\\
18479$	-0005		$	&	1	&		&		&		&		&		&	1	&		&		&		&		&		&		&	7	&	0.0407 	\\
18507$	+0110		$	&	1	&	1	&	1	&	1	&	1	&	1	&	1	&		&	1	&	1	&	1	&	1	&	1	&	19	&	0.5367 	\\
18507$	+0121		$	&	1	&	1	&	1	&	1	&	1	&	1	&	1	&		&	1	&	1	&	1	&	1	&	1	&	19	&	0.2569 	\\
18517$	+0437		$	&	1	&		&		&		&	1	&	1	&	1	&		&		&		&	1	&	1	&	1	&	13	&	0.0430 	\\
19078$	+0901	c1	$	&	1	&		&		&		&		&		&		&		&		&		&		&		&	1	&	6	&	0.0401 	\\
19078$	+0901	c2	$	&	1	&		&		&		&		&		&	1	&		&		&		&		&	1	&	1	&	9	&	0.1126 	\\
19078$	+0901	c3	$	&	1	&		&		&		&		&		&	1	&		&		&		&		&		&		&	7	&	0.0094 	\\
19095$	+0930		$	&	1	&		&		&		&		&		&	1	&		&		&		&	1	&	1	&	1	&	10	&	0.0712 	\\
Total		&	83	&	20	&	24	&	13	&	35	&	35	&	93	&	6	&	6	&	22	&	42	&	70	&	87	&	770	&		\\
\end{longtable}
\begin{flushleft}
\textbf{Notes:} a: The number of transitions, as shown in Table \ref{Table 1}, indicates that some emission lines are blends of two transitions. 
F: As described in Section \ref{obs}, the issue with the window frequency settings resulted in the absence of this emission line in these sources. Core marked with “x” indicate sources with no COM lines above 3$\sigma$, which are therefore not classified as hot cores in this work.
\end{flushleft}
\end{longrotatetable}

\section{Spectra and XCLASS Fittings} \label{app:C}

\begin{figure*}[ht!]
\centering
\includegraphics[width=\linewidth]{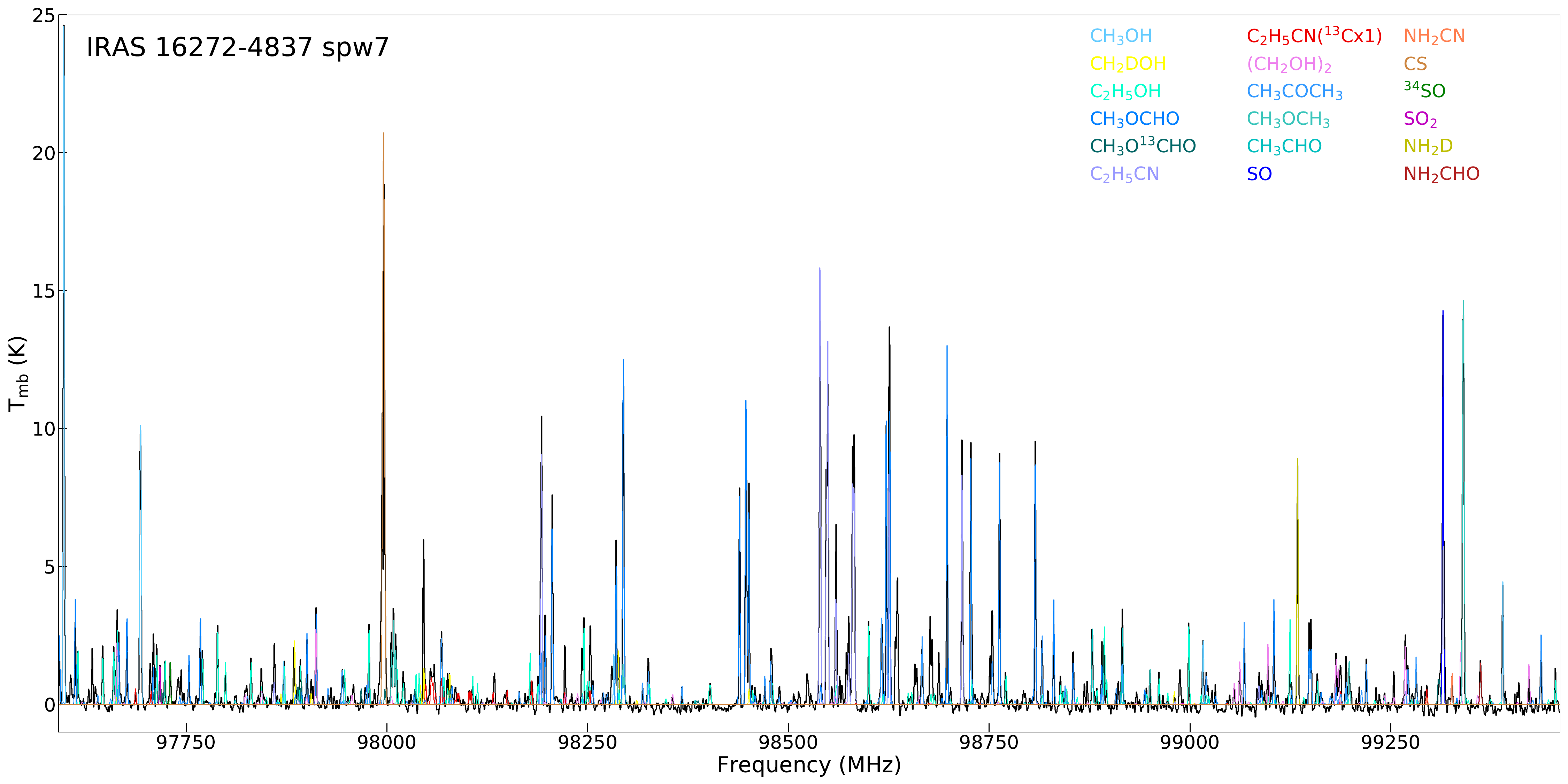}
\vskip 0.8cm
\includegraphics[width=\linewidth]{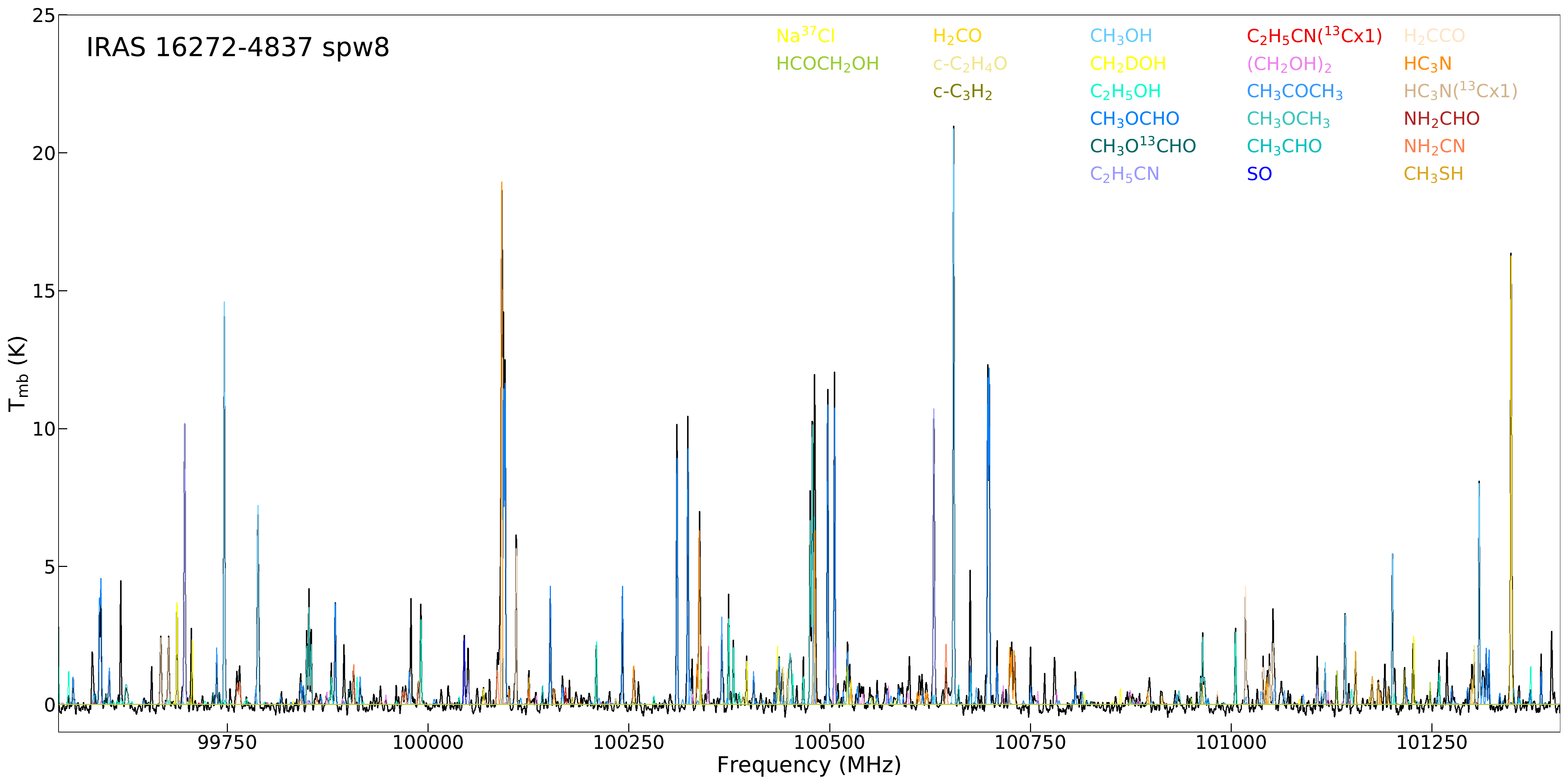}
\caption{%
In our previous work, more than 30 molecular emission lines have been securely identified toward the chemically rich hot core IRAS 16272-4837c1, confirming the reliability of the transitions used in the present study.
}
\label{fig:16272c1}
\end{figure*}

\begin{figure}[htbp]
\centering
\includegraphics[width=0.45\textwidth]{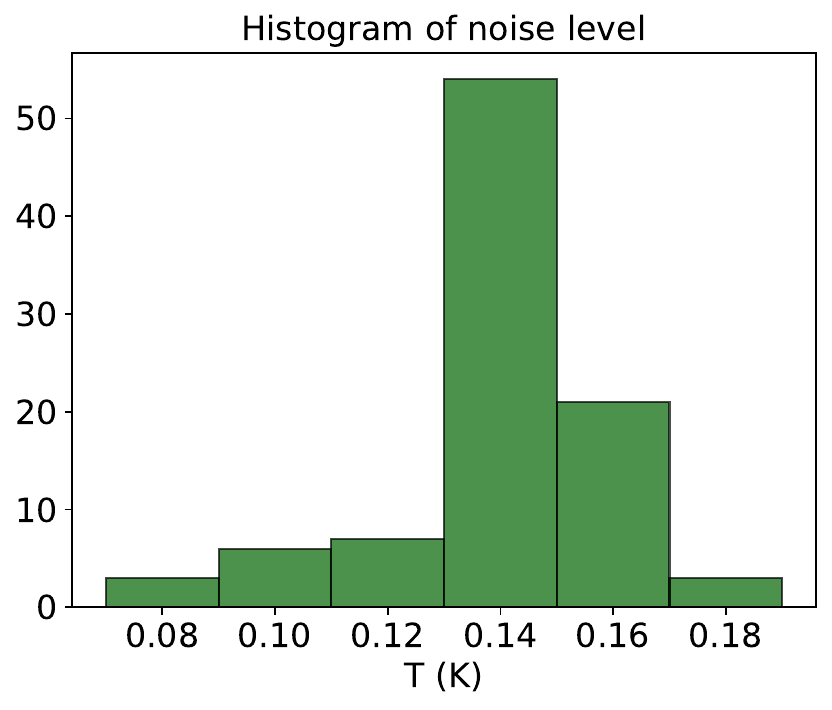}
\caption{1$\sigma$ noise-level (K) statistics of the sample, SPW 7 and SPW 8.}
\label{fig:noise}
\end{figure}

Fig.~\ref{fig:16272c1} presents the molecular-line spectrum (average within one beam of IRAS 16272-4837 c1) that we identified in ATOMS 3 mm SPWs 7 and 8 in our previous, still-unpublished work. Fig.~\ref{fig:noise} shows the noise distribution for our full sample of 94 cores; the vast majority cluster around 0.14 K.

\subsection{A/E Symmetry Species in the XCLASS Formalism}

In the XCLASS framework, the separation of A- and E-type methanol is implemented at the database level. We construct and employ a dedicated SQLite database file (.db) compatible with XCLASS, which is exported exclusively from the CDMS. No spectroscopic information from other catalogs is used in this work. Within this database, A- and E-type methanol are defined as two independent species with distinct species names, each associated with its own transition list and partition function table.

Under the LTE assumption, the optical depth of a transition $t$ of species $m$ is computed in XCLASS as \citep[Eq.~8 of][]{2017A&A...598A...7M}

\begin{equation}
\begin{split}
\tau_{l}(\nu) =
\sum_{t}
\frac{c^2}{8\pi \nu_t^2}
A_{ul}^{(t)}
\, N_{\mathrm{tot}}
\, \frac{g_u^{(t)} e^{-E_l^{(t)}/kT_{\mathrm{ex}}}}
{Q_m(T_{\mathrm{ex}})}
\\
\times
\left(1 - e^{-h\nu_t/kT_{\mathrm{ex}}}\right)
\phi_t(\nu)
\end{split}
\end{equation}

where the upper-state degeneracy $g_u$ and the partition function $Q_m(T_{\mathrm{ex}})$ are retrieved from the database for the corresponding species.

Because partition functions in XCLASS are defined and accessed on a per-species basis, the database-level separation of A- and E-type methanol necessarily requires independent partition functions, $Q(A)$ and $Q(E)$. The use of a merged partition function after species separation would lead to an inconsistent normalization of level populations through the factor $g_u/Q$.

The partition function for each symmetry species is computed following the standard definition
\begin{equation}
Q_s(T) = \sum_i g_i^{(s)} \exp\!\left(-\frac{E_i^{(s)}}{kT}\right), \qquad s \in \{A, E\},
\end{equation}
where the summation includes only energy levels belonging to the corresponding symmetry manifold.

For the E-type species, which exhibits an intrinsic twofold symmetry-related degeneracy, this degeneracy is accounted for consistently between the transition catalog and the partition function. If the degeneracy is not explicitly included in the level degeneracies $g_i$ provided by the CDMS transition list, it is incorporated in the computation of $Q(E)$, ensuring statistical consistency with the XCLASS formalism.

\section{Robustness and Biases of the CDR}
\label{app:D}

\subsection{Definition and empirical validation}
\label{app:D1}

The CDR provides a practical and robust proxy for molecular richness in large spectral-line surveys such as ATOMS and QUARKS. CDR is defined as the fraction of spectral channels, within the observed frequency range, that exceed a fixed SNR threshold. By construction, CDR requires no line identification, profile decomposition, or manual tuning of detection thresholds, making it particularly suitable for homogeneous statistical analyses.

We empirically validated CDR using 94 methanol \vt~hot-core candidates (Fig.~\ref{fig:cdrs}). 
Sources previously reported by \citet{2022MNRAS.511.3463Q} are marked with circles ($\circ$), while newly identified candidates are shown as crosses ($\times$). 
Objects hosting hydrogen recombination lines (primarily \ha) are highlighted in red \citep{2025arXiv251101285M}. 
Fig.~\ref{fig:cdrs} compares several refined indicators against the ``zero-prior'' CDR (see figure caption for details), demonstrating that CDR captures molecular richness in a manner broadly consistent with more traditional metrics.

Because line broadening artificially inflates the number of channels above a fixed SNR threshold, we apply a linewidth normalization:
\begin{equation}
\mathrm{CDR}_{\mathrm{norm}} = 
\mathrm{CDR}_{\mathrm{Core}} \times 
\frac{6.0~\mathrm{km~s^{-1}}}{\Delta v_{\mathrm{meth}}},
\end{equation}
where $\Delta v_{\mathrm{meth}}$ is the source-specific methanol linewidth.
After normalization, the 1-$\sigma$ scatter is
$\lg \mathrm{CDR}_{\mathrm{norm}} = -2.00 \pm 0.22$, 
corresponding to 
$\lg \mathrm{CDR} = -1.85 \pm 0.20$ in the 3\,mm band.
Based on this distribution, we adopt
\begin{equation}
\mathrm{CDR}_{\mathrm{norm}} > 1.7\% \quad \text{or} \quad
\mathrm{CDR} > 2.2\%
\end{equation}
as a \emph{sufficient} (but not necessary) criterion for identifying hot cores at 3\,mm.

\subsection{CDR and line-intensity bias}
\label{app:D2}

A potential concern is whether CDR is biased by absolute line intensity, for example due to beam dilution, lower $N_{\mathrm{H_2}}$, or intrinsically lower molecular column densities.
To quantify this effect, we consider a Gaussian line profile clipped at $y = 1\,\mathrm{K}$, corresponding to our adopted 3$\sigma$ detection threshold.
For two cores whose \emph{same} transition has peak amplitudes $A$ and $B$ (both $\gg 3\sigma$), the ratio of the velocity widths that survive the clip is
\[
\frac{\Delta v_{A}}{\Delta v_{B}}
= \sqrt{\frac{\ln A}{\ln B}}.
\]

When both $A$ and $B$ are well above the noise ($A, B \gg 1\,\mathrm{K}$), their logarithms differ only weakly, and the width ratio approaches unity. 
In this regime, differences in total line intensity have a negligible impact on CDR.
Only when a line hovers close to the 3$\sigma$ threshold does clipping remove the weak wings, producing an apparent loss of detectable channels.

Importantly, this is precisely the regime in which full spectral fitting methods (e.g.\ XCLASS, \texttt{spectuner}, or other LTE-based approaches) also fail to converge or systematically under-estimate molecular content.
CDR therefore carries no intrinsic disadvantage relative to more sophisticated, but equally SNR-limited, techniques.

\subsection{Limitations and interpretation}
\label{app:D3}

Within a finite beam, low-mass or very distant sources may remain unresolved, leading to intrinsically low beam-averaged column densities or severe beam dilution (low filling factors). 
In such cases, spectral lines may fall below the \emph{current} 3$\sigma$ sensitivity limit, reducing the reliability of any chemical richness metric, including CDR, particularly for sources at distances $\gtrsim 10$~kpc.

This behaviour should be interpreted as an \emph{under-estimate} imposed by the present data quality rather than a failure unique to CDR. 
We therefore adopt a uniform 3$\sigma$ criterion instead of a fixed brightness temperature (e.g.\ 0.3\,K) or an absolute column-density threshold: emission undetected at the current noise level is treated as ``chemically poor'' \emph{in this data set}. 
Should higher-sensitivity or higher-resolution observations recover these lines, such changes reflect the improved data rather than a revision of the diagnostic itself.

Sources falling below the adopted CDR thresholds are thus natural targets for follow-up observations, either with higher-resolution 3\,mm imaging within the ongoing ATOMS program or by targeting intrinsically stronger transitions (larger Einstein-$A$ coefficients) in ALMA Bands~6/7 to retrieve the missing molecular signal.

\begin{figure}[htbp]
\centering
\includegraphics[width=0.45\textwidth]{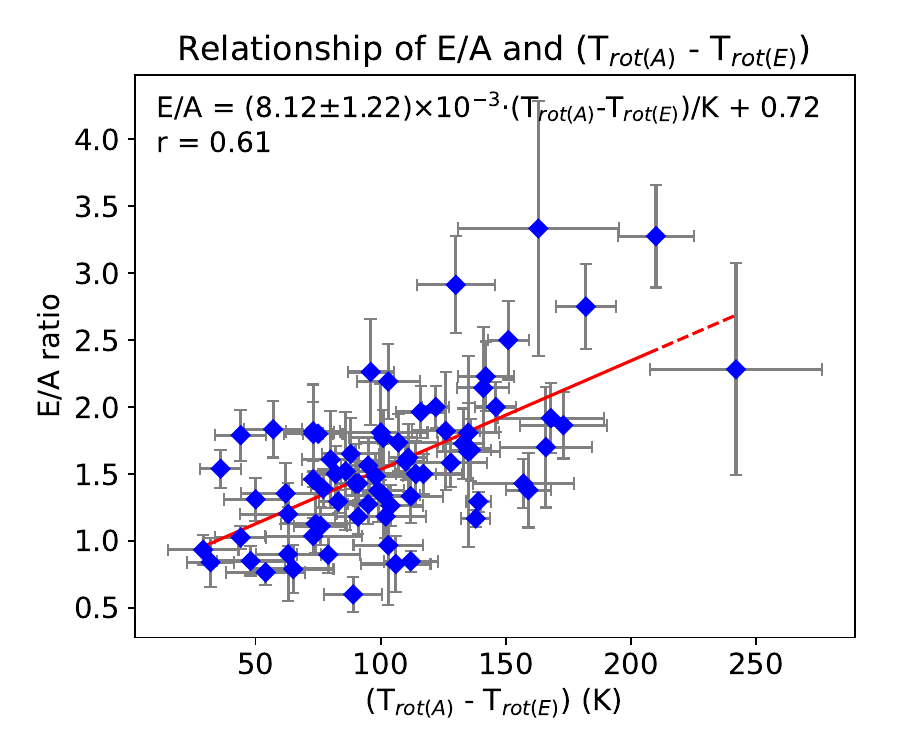}
\caption{The E/A ratio versus the rotational temperature difference. When the rotation temperature of \eme\ is underestimated, its column density is correspondingly overestimated.}
\label{fig:EA_TA-TE}
\end{figure}


\begin{figure*}[htbp]
\centering
\setcounter{figure}{10}
\includegraphics[width=0.95\textwidth]{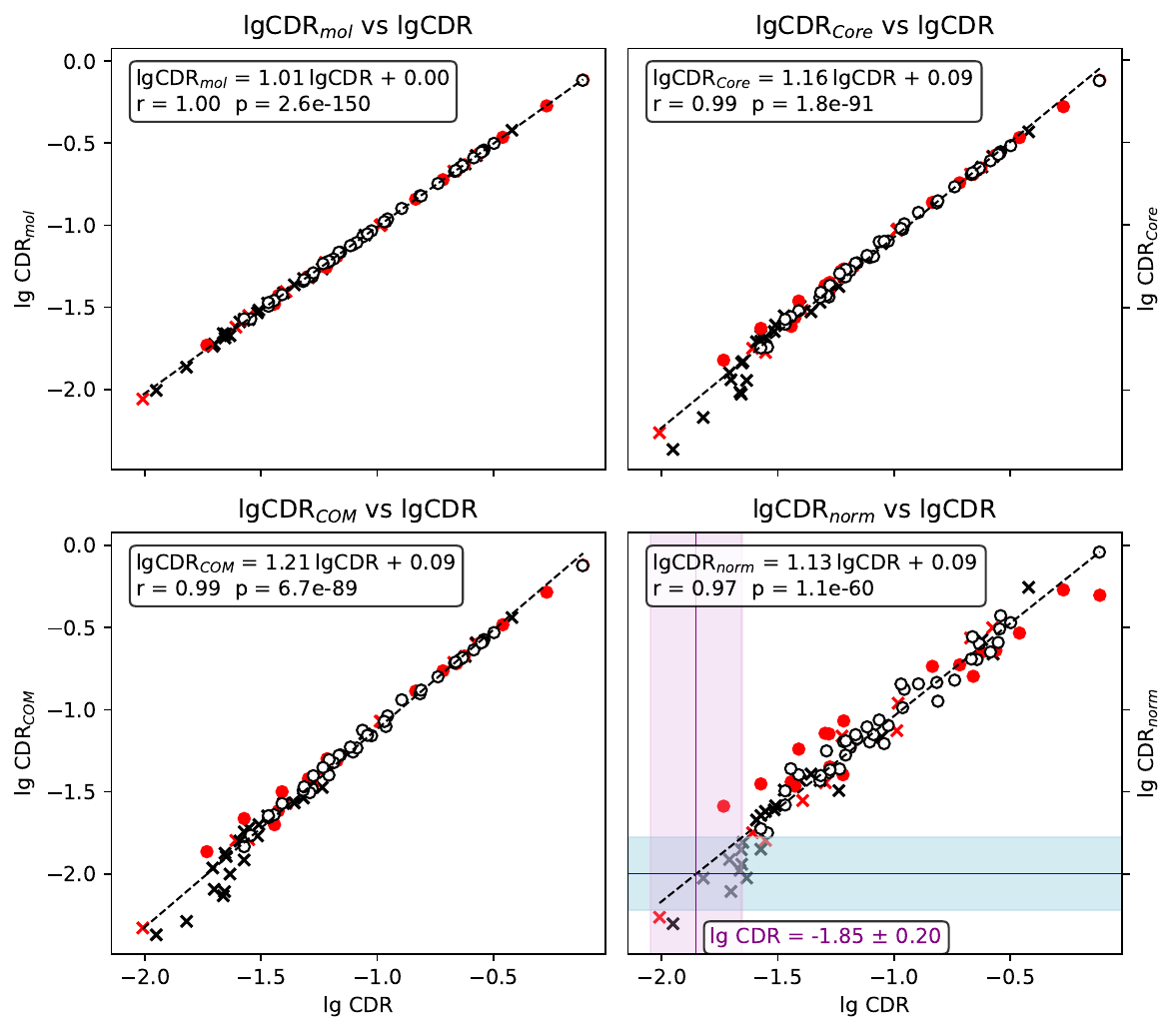}
\caption{
We compared the differences among various CDR extraction methods.
The horizontal axis shows CDR: the signal fraction over the entire spectral window, with no channels excluded.
Top-left, CDR$_{mol}$: channels associated with several radio recombination lines (primarily H40$\alpha$) are excluded, identical to the channels used for $F_{\mathrm{mol}}$ in Section~\ref{sec 4.2}.
Top-right, CDR$_{Core}$: channels near molecular transitions detected outside the hot core are excluded, following Section~\ref{sec 3.3}.
Bottom-left, CDR$_{COM}$: all channels not belonging to COM molecules are excluded.
Bottom-right, CDR$_{norm}$: normalized version of CDR$_{Core}$.
Symbols ``o'' mark hot cores reported in \citet{2022MNRAS.511.3463Q}, while ``x'' indicates new hot core candidates identified in this work; red denotes sources associated with hydrogen recombination lines.
The blue line and shaded band indicate CDR$_{norm}$=0.01 and the uncertainty propagated from the sample velocity distribution; their projection onto the CDR axis is shown by the purple line and band.
In our method, sources falling above and to the right of this band are identified as hot cores.
}

\label{fig:cdrs}
\end{figure*}

\clearpage
\onecolumngrid
\bibliography{MeOH}{}
\bibliographystyle{aasjournalv7}



\end{document}